\documentclass[11pt,english]{article}
\usepackage[T1]{fontenc}
\usepackage[latin9]{inputenc}
\usepackage[a4paper]{geometry}
\usepackage{float}
\usepackage{amsmath}
\usepackage{amsthm}
\usepackage{amssymb}
\usepackage{graphicx}
\usepackage{setspace}
\usepackage{color}
\makeatletter

\allowdisplaybreaks

\providecommand{\tabularnewline}{\\}
\newcommand{\lyxdot}{.}

\theoremstyle{remark}
\newtheorem{rem}{\protect\remarkname}
\theoremstyle{plain}
\newtheorem{lem}{\protect\lemmaname}
\theoremstyle{plain}
\newtheorem{thm}{\protect\theoremname}

\usepackage{babel}
\usepackage{pdflscape}

\makeatother

\usepackage{babel}
\providecommand{\lemmaname}{Lemma}
\providecommand{\remarkname}{Remark}
\providecommand{\theoremname}{Theorem}

\newcommand{\no}{\nonumber}

\newcommand{\ep}{\epsilon}
\newcommand{\la}{\label}
\newcommand{\be}{\begin{eqnarray}}
\newcommand{\ee}{\end{eqnarray}}
\newcommand{\bestar}{\begin{eqnarray*}}
\newcommand{\eestar}{\end{eqnarray*}}
\newcommand{\lam}{\lambda}

\newcommand{\al}{\alpha}
\newcommand{\de}{\delta}

\newcommand{\ga}{\gamma}

\newcommand{\ignore}[1]{}
{} \theoremstyle{plain}

\def\d{\delta}

\def\>{\geq}
\def\<{\leq}

\begin{document}
\title{LOCALLY TRIMMED LEAST SQUARES: CONVENTIONAL INFERENCE IN POSSIBLY
NONSTATIONARY MODELS}
\author{Zhishui Hu\thanks{\baselineskip=0.4 true cm International Institute
of Finance, School of Management, University of Science and Technology of
China, Hefei, Anhui 230026, China, email: huzs@ustc.edu.cn.},  Ioannis Kasparis\thanks{\baselineskip=0.4
true cmUniversity of Cyprus, Nicosia, Cyprus, email: kasparis@ucy.ac.cy.} and Qiying Wang\thanks{\baselineskip=0.4 true cmSchool of Mathematics
and Statistics, The University of Sydney, NSW 2006, Australia, email:
qiying@maths.usyd.edu.au.} }
\maketitle
\begin{abstract}
A novel IV estimation method, that we term \textit{Locally Trimmed
LS (LTLS)}, is developed which yields estimators with (mixed) Gaussian
limit distributions in situations where the data may be weakly or
strongly persistent. In particular, we allow for nonlinear predictive
type of regressions where the regressor can be stationary short/long
memory as well as nonstationary long memory process or a nearly integrated
array. The resultant t-tests have conventional limit distributions
(i.e. $N(0,1)$) free of (near to unity and long memory) nuisance
parameters. In the case where the regressor is a fractional process,
no preliminary estimator for the memory parameter is required. Therefore,
the practitioner can conduct inference while being agnostic about
the exact dependence structure in the data. The LTLS estimator is
obtained by applying certain \textit{chronological trimming} to the
OLS instrument via the utilisation of appropriate kernel functions
of time trend variables. The finite sample performance of LTLS based
t-tests is investigated with the aid of a simulation experiment. An
empirical application to the predictability of stock returns is also
provided.
\end{abstract}

\section{Introduction}

It is well known that under nonstationarity   regression estimators
do not have conventional limit distributions in general. As a consequence,
the inferential procedures developed for stationary data are not applicable
under nonstationarity. A number of early studies in the area of nonstationary
econometrics (e.g. Phillips and Hansen, 1990; Johansen, 1995; Phillips,
1995; Robinson and Hualde 2003) develop inferential procedures suitable
for nonstationary models, however these methods are not valid in general
under stationarity. In fact, it is well known that methods such as
FMLS (c.f. Phillips, 1995) may exhibit severe size distortions even
under local deviations from the (fractional) unit root paradigm. This
duality in inference, has made empirical work in time series econometrics
elusive. Practitioners typically need to make preliminary (some times
ad hoc) assumptions about the persistence level in the data or apply
some sort of pre-testing -and therefore expose inference to problems
associated to pre-testing- before proceeding to estimation and inference.
A number of studies has attempted to address this issue using conservative
confidence intervals (for a review see Mikusheva (2007), Phillips
(2014) and the references therein). The more recent work of Magdalinos
and Phillips (2009; MP hereafter) (see also Kostakis, Magdalinos and
Stamatogiannis (2015) for refinements and additional results) follows
a completely different direction. MP propose an IV estimator (IVX)
that that has mixed Gaussian limit distribution at the expense of
an arbitrary reduction in the convergence rate, relative to that of
the OLS estimator.

In this paper we follow an approach similar to the pioneering work
of MP. To fix ideas consider the simple model
\begin{equation}
y_{k}=\beta x_{k}+u_{k},\text{ }k=1,...,n,\label{model0}
\end{equation}
where $x_{k}$ is a nearly integrated (NI) and $x_{k}$ predetermined
with respect to some martingale difference error term ($u_{k}$).\ MP
construct the so called IVX instrument by applying the following linear
filtering to the OLS instrument ($x_{k}$)
\begin{equation}
Z_{kn}=\sum_{j=0}^{k-1}\left(1+\frac{c_{z}}{n^{b}}\right)^{j}(x_{k-j}-x_{k-j-1})\text{, }\label{IVX}
\end{equation}
for some $c_{z}<0$ and $0<b<1$. This linear filtering transforms
$x_{k}$ into a mildly integrated process (e.g. see Giraitis and Phillips,
2006; Phillips and Magdalinos, 2007) that is less persistent than
a NI array (e.g. $x_{k}$). By choosing $b$ arbitrary close to unity,
the\ reduction in the signal of the instrument results into an arbitrary
small reduction in the convergence rate of the IVX estimator, relative
to that of the OLS, and this is sufficient for a martingale CLT to
operate, rendering IVX based inference conventional. The choice of
$b$ is important to inference with smaller $b$\
resulting in better size control at the expense of asymptotic power.
Note that as $b\uparrow1$, $Z_{kn}$\ approximates the NI process
$x_{k}$ and the IVX estimator resembles the behaviour of the OLS
estimator. The recent work of Yang, Long, Peng and Cai (2019) generalises
the IVX method to regression models with serially correlated regression  (parametric AR)
errors, whilst Demetrescu, Georgiev, Rodrigues and Taylor (2020) apply
a modified version of the IVX estimator to test for episodic predictability
in stock returns.

We consider an alternative method for reducing the signal of the OLS
instrument. Let $K$ be an integrable kernel function and set
\[
Z_{kn}=K\left[c_{n}\left(k/n-\tau\right)\right]x_{k}\text{,}
\]
where $c_{n}$ is a positive deterministic sequence such that$\ c_{n}^{-1}+c_{n}n^{-1}\rightarrow0$
and $0<\tau<1$. For simplicity set $\tau=1/2$ and $K(0)=1$. In
this case the kernel function extracts information from the OLS instrument
for observations near the middle of the sample. In particular, $Z_{kn}\approx x_{k}$
i.e. when $k\approx n/2$, and $Z_{kn}\approx0$\ when $k$ is far
from $n/2$. In other words certain \textit{chronological trimming}
applies around the ``\textit{chronological point} $\tau$''. By
allowing the $c_{n}$\
sequence to diverge at an arbitrary slow rate, the resultant IV (LTLS)
estimator attains an arbitrary slower convergence rate relative to
the OLS estimator. In principle, it is possible to extract information
around multiple chronological points $0<\tau_{1}<...<\tau_{l_{n}}<1$
where $l_{n}$\,is either fixed or $l_{n}\rightarrow\infty$ such
that $l_{n}=o(c_{n})$.
In this case the relevant instrument is
\begin{equation}
Z_{kn}=\sum_{j=1}^{l_{n}}K\left[c_{n}\left(k/n-\tau_{j}\right)\right]x_{k}.\label{LTLS-Z}
\end{equation}
As long as the LTLS estimator converges at slower rate, than the OLS
estimator, limit theory is mixed Gaussian for nonstationary regressor
covariates and Gaussian for stationary. In particular, the reduction
in the signal of the OLS instrument allows a martingale CLT (c.f.
Wang, 2014) to operate even if $x_{k}$ is nonstationary. Notice that
if $c_{n}$ is too small or if too many chronological points ($l$)
are employed, then $Z_{kn}$ approximates the OLS instrument and as
a consequence LTLS based inference resembles OLS based inference.
This can be easily seen, if a vanishing sequence $c_{n}$ is employed.
Note that for $c_{n}\rightarrow0$, $Z_{kn}\approx l_{n}K(0)x_{k}$.

Our theoretical framework allows for a wide range of stationary and
nonstationary linear processes as well NI arrays. In particular, $x_{k}$
can be a stationary or a nonstationary fractional process. Consider
the LTLS estimator of $\beta$ in (\ref{model0}) that utilises the
instrument of (\ref{LTLS-Z}) i.e. $\hat{\beta}=\sum_{k=1}^{n}Z_{kn}y_{k}/\sum_{k=1}^{n}Z_{kn}x_{k}$.
Let $t\in\lbrack0,1]$ and suppose that $x_{k}$ is a nonstationary
process such that for some $d_{n}\rightarrow\infty$, $d_{n}^{-1}x_{\lfloor nt\rfloor}\Rightarrow X_{t}$
in $D[0,1]$\ where $X_{t}$ is a continuous process. For instance
$X_{t}$ can be a fractional BM or a fractional Ornstein-Uhlenbeck
process (see Remark \ref{remark2}\ below) depending on some memory
or near-to-unity nuisance parameter.\ Then we have
\[
d_{n}\sqrt{\frac{nl_{n}}{c_{n}}}\left(\hat{\beta}-\beta\right)\rightarrow_{d}\mathbf {MN}\left(0,E\left(u_{t}^{2}\right)\frac{\int_{%TCIMACRO{\U{211d}}%
%BeginExpansion
\mathbb{R}%EndExpansion
}K^{2}(x)dx}{\left(\int_{%TCIMACRO{\U{211d}}%
%BeginExpansion
\mathbb{R}%EndExpansion
}K(x)dx\right)^{2}\int_{0}^{1}X_{t}^{2}dt}\right).
\]
Because $c_{n}\rightarrow\infty$ and $l_{n}=o(c_{n})$ the convergence
rate of the LTLS is slower than that of the OLS estimator ($d_{n}\sqrt{n}$).
Further, note that nuisance parameters affect the limit distribution
only via the mixing variate $\left[\int_{0}^{1}X_{t}^{2}dt\right]^{-1}$
and as a consequence the studentised LTLS estimator has standard normal
limit distribution. Interestingly, the limit variance shown above
is the same, up to a constant, to that of the FMLS estimator for the
case where $x_{k}\sim I(1)$.

We mention that the constant that features in the limit variance
of the LTLS estimator above, can be made arbitrarily small by an appropriate
choice of the kernel function. For example suppose that $K(x)=\left(2\pi\varsigma^{2}\right)^{-1/2}\exp\left(-\frac{x^{2}}{2\varsigma^{2}}\right)$.
Then
\[
E\left(u_{t}^{2}\right)\int_{%TCIMACRO{\U{211d}}%
%BeginExpansion
\mathbb{R}%EndExpansion
}K^{2}(x)dx\Big/\left(\int_{%TCIMACRO{\U{211d}}%
%BeginExpansion
\mathbb{R}%EndExpansion
}K(x)dx\right)^{2}=\frac{E\left(u_{t}^{2}\right)}{2\sqrt{\pi\varsigma^{2}}}\rightarrow0
\]
as $\varsigma^{2}\rightarrow\infty$.\footnote{
\begin{eqnarray*}
\int_{%TCIMACRO{\U{211d}}%
%BeginExpansion
\mathbb{R}%EndExpansion
}K^{2}(x)dx&=&\left(2\pi\varsigma^{2}\right)^{-1}\int_{\mathbb{R}}\exp\left(-\frac{x^{2}}{2(\varsigma^{2}/2)}\right)dx
=\left(2\pi\varsigma^{2}\right)^{-1}\sqrt{2\pi(\varsigma^{2}/2)}
=\frac{1}{2\sqrt{\pi\varsigma^{2}}}.
\end{eqnarray*}
} Nevertheless, choosing a large value of the kernel variance parameter
has the same effect as choosing a small value for $c_{n}$. Therefore
as $\varsigma^{2}\rightarrow\infty$, the LTLS estimator approximates
the OLS estimator.

It should be further noted that for nonstationary fractional covariates (i.e.
$I(d)$, $d>1/2$), methods like FMLS (e.g. Phillips, 1995) or the
spectral GLS of Robinson and Hualde (2003) (see also Hualde and Robinson,
2010) are asymptotically equivalent the Gaussian pseudo maximum likelihood
and therefore asymptotically efficient (c.f. Phillips, 1991). The
key feature of these methods is to induce asymptotically mixed Gaussian
estimators by certain \textit{modification} in the dependent variable
that involves (fractionally) differencing the covariates. In the context
of (\ref{model0}) such differencing takes the form $(I-L)^{\hat{d}}x_{k}$,
where $L$ is the lag operator and$\ \hat{d}$\ is a preliminary
estimator for the memory parameter of $x_{k}$. Nevertheless, if there
is a local deviation (order $O(n^{-1})$) from the (fractional) unit
root model, the aforementioned methods yield mixed Gaussian limit
theory only if the following quasi fractional differencing is applied
\[
\left(I-\left(c/n\right)L\right)^{\hat{d}}x_{k},
\]
where $c$ is a local to unity parameter. A non trivial value for
the local to unity parameter however renders the aforementioned methods
infeasible because of the lack of identifiability of $c$. It is well
known that if $c\neq0$, inference based on methods like FMLS are
prone to severe size distortions even if there is moderate correlation
between the regressor and the regression error.

The remaining of this work is organised as follows. Section 2 provides
basic limit theory for locally trimmed functionals of stationary and
nonstationary processes. This limit theory is utilised in Section
3 for exploring the limit properties of the LTLS estimation and inference.
Section 4 provides a simulation study and Section 5 an empirical application
on the predictability of stock returns.

Throughout this paper we make use of the following  notation. For two
deterministic sequences $a_{n}$ and $b_{n}$, $a_{n}\sim b_{n}$
denotes $\lim_{n\rightarrow\infty}a_{n}/b_{n}=1$. $1\left\{ A\right\} $
is the indicator function on set $A$. We may write the integral $\int_{%TCIMACRO{\U{211d}}%
%BeginExpansion
\mathbb{R}%EndExpansion
}f(x)dx$ as $\int f$. $\Rightarrow$ denotes weak convergence in the space
$D[0,1]$. For a vector $x$, $\left\Vert x\right\Vert $\ is its
inner product norm and $x^{\prime}$ its transpose. By $[x]$
we denote the integer part of a positive number $x$. Finally, diag$\{a_{1},...,a_{p}\}$
denotes a $p\times p$ diagonal matrix with elements $\{a_{1},...,a_{p}\}$
on the main diagonal, $\to_d$ denotes the convergence in distribution and $Y:=\mathbf{MN}(\mathbf{0}, \sum)$ denotes a Gaussian variate (mixing normal) with characteristic function $f(t)=E e^{it' Y}=Ee^{-t'\sum t/2}$.

\section{Asymptotics for locally trimmed sample functionals}

In this section we develop basic limit theory for locally trimmed
(LT) sample functionals of stationary and nonstationary processes.
Our basic limit theory is utilised in Section 3 for the asymptotic analysis
of the LTLS estimator.\ Let $\left\{ x_{k}\right\} _{1\leq k\leq n}$
be a scalar time series process and $\{X_{nk}\}_{1\leq k\leq n,n\geq1}$
be some scalar random array. Further, let $K
$ be an integrable kernel function and $\ g(.)=\left[g_{1}(.),...,g_{p}(.)\right]^{\prime}$,
where, for each $i=1,...,p$, $g_{i}
$ is a measurable function. For $l\in
\mathbb{N}
$ and$\ 0<\tau_{1}<...<\tau_{l}<1$, set
\begin{eqnarray*}
S_{1n,l}&=&\frac{c_{n}}{n}\sum_{k=1}^{n}g(x_{k})\left\{ \frac{1}{l}\sum_{j=1}^{l}K\left[c_{n}(k/n-\tau_{j})\right]\right\} ,
\\
M_{1n,l}&=&\sqrt{\frac{c_{n}}{n}}\sum_{k=1}^{n}g(x_{k})\left\{ \frac{1}{\sqrt{l}}\sum_{j=1}^{l}K\left[c_{n}(k/n-\tau_{j})\right]\right\} u_{k},
\\
S_{2n,l}&=&\frac{c_{n}}{n}\sum_{k=1}^{n}g(X_{nk})\left\{ \frac{1}{l}\sum_{j=1}^{l}K\left[c_{n}(k/n-\tau_{j})\right]\right\} ,
\\
M_{2n,l}&=&\sqrt{\frac{c_{n}}{n}}\sum_{k=1}^{n}g(X_{nk})\left\{ \frac{1}{\sqrt{l}}\sum_{j=1}^{l}K\left[c_{n}(k/n-\tau_{j})\right]\right\} u_{k},
\end{eqnarray*}
where $c_{n}$ is a sequence of positive constants, $l$ either fixed
or $l\rightarrow\infty$ as $n\rightarrow\infty$, and $u_{k}$ together
with an appropriate filtration $\left\{ \mathcal{F}_{k}\right\} $
forms a martingale difference sequence (such that $X_{nk}$, $x_{k}$
are $\mathcal{F}_{k-1}$-measurable). The limit theory of the LTLS
estimator relies on the asymptotics of $\left\{ S_{jn,l},M_{jn,l}\right\} _{j=1}^{2}$.
Limit theory for the functionals $\left\{ S_{1n,l},M_{1n,l}\right\} $
is relevant for stationary  regressors whilst $\left\{ S_{2n,l},M_{2n,l}\right\} $
for nonstationary. In fact, it is assumed that $X_{nk}$ satisfies
some FCLT.\ The term $S_{2n,l}$ resembles certain functionals considered
by Phillips, Li and Gao (2017) who study the estimation of cointegrated
models with smooth time varying parameters (TVP). The aforementioned
work considers terms of the form
\begin{eqnarray*}
\frac{c_{n}}{n}\sum_{k=1}^{n}X_{nk}^{2}K\left[c_{n}(k/n-\tau)\right],\text{ }0<\tau<1,
\end{eqnarray*}
where $X_{nk}$\ is an $I(1)$ process\ normalised by $\sqrt{n}$.\ As
explained below, under our assumptions $X_{nk}$\ can be an appropriately
normalised $I(d)$, $d>1/2$, process or a NI array (possibly driven
by fractional errors). Therefore the limit results provided in this
section are also relevant to the estimation of TVP models for the
case where the covariate is a general nonstationary process satisfying
some FCLT (see Assumption \textbf{A3} below).\

\medskip
 To facilitate  basic
limit results, we make use of the following  conditions.

\medskip

\begin{itemize}
\item[ \textbf{A1 }] (innovations):
 $\left\{ \eta_{k},\mathcal{F}_{k}\right\} _{k\geq1}$, where $\eta_{k}^{\prime}=\left(\xi_{k+1},u_{k}\right)$ and
$\mathcal{F}_{k}=\sigma(u_{k},u_{k-1},...,u_{1};\xi_{j},j\leq k+1),$
forms a $2$-dimensional martingale difference satisfying the following
conditions:
\begin{itemize}
\item[(a)] $\sup_{k\geq1}E(u_{k}^{2}I(|u_{k}|\geq M)|\mathcal{F}_{k-1})=o_{P}(1),$
as $M\rightarrow\infty$;
\item[(b)] $\sup_{k\geq1}E(\xi_{k}^{2}I(|\xi_{k}|\geq M)|\mathcal{F}_{k-1})=o_{P}(1),$
as $M\rightarrow\infty$;
\item[(c)] there exists a positive definite matrix:
\[
\Sigma=\left[\begin{array}{cc}
\sigma_{\xi}^{2} & \sigma_{\xi u}\\
\sigma_{u\xi} & \sigma_{u}^{2}
\end{array}\right]
\]
so that, for all $k\geq1$, $E\left({\eta}_{k}{\eta}_{k}^{\prime}\mid\mathcal{F}_{k-1}\right)=\Sigma,\ a.s.$
\end{itemize}
\item[\textbf{A2}] (stationary process): $x_{k}$ is an ergodic (strictly) stationary random sequence
and a functional of $\xi_{k},\xi_{k-1},...$ satisfying that $E\left\Vert g(x_{k})\right\Vert ^{2+\delta}<\infty$
for some $\delta>0$.
\item[\textbf{A3}] (nonstationary process and invariance principle): $X_{nk}=d_n^{-1}x_{k}$ , where $0<d_n^2= var(x_n)\to\infty$ and $x_k$ is a functional of $\xi_{k},\xi_{k-1},...$ (depending on $n$ is allowed) so that, on  $D_{%
%TCIMACRO{\U{211d} }%
%BeginExpansion
\mathbb{R}
%EndExpansion
^{3}}[0,1]$,
\be
\left[\frac{1}{\sqrt{n}}\sum_{k=1}^{[nt]}\xi_{k},\ \frac{1}{\sqrt{n}}\sum_{k=1}^{[nt]}\xi_{-k},\ X_{n,[nt]}\right]\Rightarrow\left[B_{1t},B_{2t},X_{t}\right], \label {addmain}
\ee
where $B_{1t}$ and $B_{2t}$ are two independent Gaussian process
with mean zero and stationary independent increments, and $X_{t}$
is a continuous process that depends only on functionals of $\{B_{1t}\}_{0\leq t\leq1}$
and $\{B_{2t}\}_{0\leq t\leq1}$.
%\begin{itemize}

\item[\textbf{A4}] (kernel function and restrictions on $\tau_j, l_n$ and $c_n$):
\begin{itemize}
\item[(a)] $K(x)$ is a positive real function having a compact support;
\item[(b)] $0<c_{n}\rightarrow\infty$ and $c_{n}/n\rightarrow0$;
\item[(c)]  $\tau_{j}=j/(l_{n}+1)$ where  $j=1,...,l_{n}$  with $l_{n}^{-1}+c_{n}^{-1}l_{n}\rightarrow0$.
\end{itemize}

\end{itemize}

We remark that the innovation process $\left\{ \mathbf{
\eta }_{k},\mathcal{F}_{k}\right\}_{k\ge 1} $ used in  {\bf A1} is standard in literature so that both $M_{1n, l}$ and $M_{2n, l}$ have a martingale structure.  The uniform integrability conditions (a) and (b) are weak in comparison with the high moments used in previous works.  See, for instance, Wang (2014) and Wang and Phillips (2009a, b).  Since $\Sigma$ is required to be a positive definite matrix, condition (c) excludes  the  process $u_k$ to be ARCH and GARCH models. The condition (c) is  required for technical reasons, which seems to be difficult to reduce at the moment.

Stationary process given in {\bf A2} is extensively used in empirical applications where examples   include short and long memory (fractional) processes.   Typical examples on nonstationary processes  satisfying {\bf A3} have the form:
\bestar
x_{k}&=&\rho x_{k-1,}+\sum_{i=0}^{\infty }\phi _{i}\xi _{k-i},
\eestar
 where $\rho =1+c /n$ with $ c\in\mathbb{R} $
 and $\sum_{i=0}^{\infty }\phi _{i}^{2}<\infty.$ For the latter specification, (\ref {addmain}) holds with $X_t$ being a fractional Ornstein-Uhlenbeck
process. See, for instance, Buchmann and Chan (2007), Wang and Phillips (2009a, b) and Wang (2015).

As for {\bf A4}, the restriction on compact support for $K(x)$ can be relaxed if we have more conditions on $l_n$. Indeed,  in the following main results, {\bf A4} can be replaced by the following:

\begin{itemize}
\item[ $\textbf{A4}^*$] (kernel function and restrictions on $\tau_j, l_n$ and $c_n$):
\begin{itemize}
\item[(a)] $K(x)$ is an eventually
monotonic (i.e., there exists $A_1>0$ such that
$K(x)$ is monotonic on $(-\infty,-A_1)$ and $(A_1,\infty)$) positive function so that $K(x)\leq C/(1+|x|)$ and $\int K<\infty$;
\item[(b)] $0<c_{n}\rightarrow\infty$ and $c_{n}/n\rightarrow0$;
\item[(c)]  $\tau_{j}=j/(l_{n}+1)$ where  $j=1,...,l_{n}$  with $l_{n}^{-1}+c_{n}^{-1}\,l_{n}\,\log n\rightarrow0$.
\end{itemize}

\end{itemize}

We now  introduce the limit theory  for LT sample functionals.  Since  there are essential difference between $M_{1n, l}$ and $M_{2n, l}$, the main results will be presented based on stationary and nonstationary processes, separately.

\begin{thm}
\label{lemma1} Suppose  {\bf A2 } and {\bf A4} or  {\bf A4$^*$}  hold. Then, as $n\to\infty$, we have
\begin{equation}
S_{1n,l_{n}}=Eg(x_{1})\int K+o_{P}(1). \label{Lemma1a}
\end{equation}
 If in addition \textbf{A1}, then, as $n\to\infty$,
\begin{equation}
M_{1n,l_{n}}\rightarrow_{d} \mathbf{N}\left(\mathbf{0},~\sigma_{u}^{2}E\left[g(x_{1})g(x_{1})^{\prime}\right]\int K^{2}\right). \label{Lemma1b}
\end{equation}

\end{thm}

\begin{thm}
\label{lemma2}  Suppose that {\bf A3 } and {\bf A4} or  {\bf A4$^*$} hold and  $g(.)$ is continuous. Then, as $n\to\infty$, we have
\begin{equation}
S_{2n,l_{n}}=\int_{0}^{1}g(X_{n,\lfloor nt\rfloor})dt\int K+o_{P}(1)\rightarrow_{d}\int_{0}^{1}g(X_{t})dt\int K.\label{Lemma2a}
\end{equation}
If in addition {\bf A1},
jointly with (\ref{Lemma2a}), we have
\begin{equation}
M_{2n,l_{n}}\rightarrow_{d} \mathbf{MN}\left(\mathbf{0},~\sigma_{u}^{2}\int_{0}^{1}g(X_{t})g(X_{t})^{\prime}dt\int K^{2}\right).\label{Lemma2b}
\end{equation}
\end{thm}

\begin{rem}
\label{remark2}
 If we are only interesting the similar  results as those of (\ref {Lemma1a}) and (\ref {Lemma2a}), conditions {\bf A2} and {\bf A3} can be reduced. For instance, the result (\ref {Lemma2a}) still holds if only (\ref {addmain}) is replaced by $X_{n, [nt]}\Rightarrow X_t$ on $D_{\mathbb{R}}[0,1]$. See Lemma \ref {thm27} in Section 6 for more details. Furthermore, if  $x_{k}$ is a weakly nonstationary
process (i.e., $I(1/2)$ and mildly
integrated processes, where FCLTs do not apply)   as considered in  Phillips and Magdalinos (2007) and   Duffy and Kasparis (2018), some preliminary calculations suggest (see also Theorem 2.2
in Duffy and Kasparis, 2018) that
\[
\frac{c_{n}}{n}\sum_{k=1}^{n}g\big(d_n^{-1}x_{k}\big)\Big\{ \frac{1}{l_{n}}\sum_{j=1}^{l_{n}}K\left[c_{n}(k/n-\tau_{j})\right]\Big\}
\rightarrow_{d}\int_{%TCIMACRO{\U{211d}}%
%BeginExpansion
\mathbb{R}%EndExpansion
}g(x+X^{-})\varphi_{\sigma_{+}^{2}}(x)dx\int K,
\]
where $\varphi_{\sigma_{+}^{2}}(x)$ is the density of a $N\left(0,\sigma_{+}^{2}\right)$ variate
($\sigma_{+}^{2}>0$) and $X^{-}\sim N\left(0,\sigma_{-}^{2}\right)$
($\sigma_{-}^{2}\geq0$).  Discussions toward this kind of generalization, together with the investigation for trimmed
sample functionals of weakly nonstationary processes, are left   for future work.
\end{rem}

\begin{rem}
The continuity requirement in Theorem \ref {lemma2}
is not essential for (\ref{Lemma2a}) and (\ref{Lemma2b}). These
results can be extended to the case where $g$ is locally Lebesgue
integrable, if we impose more smoothness conditions on $X_{nk}$ (see for example Christopeit (2009) and the references therein).
This kind of generalisation involves more complicated derivations
and will not be pursued here in order to keep the paper under reasonable
length.

\end{rem}

\begin{rem}
\label{remark2a}
 Following the proof of Theorem \ref {lemma1}, it is easy to see that results (\ref {Lemma1a}) and (\ref {Lemma1b})  still hold if
 {\bf A4} (c) is replaced by  $\tau_{j}=j/(l+1)$ where  $j=1,...,l$,  i.e., if $l_n\equiv l$ is fixed. As for (\ref {Lemma2a}) and (\ref {Lemma2b}), if {\bf A4} (c) is replaced by  $\tau_{j}=j/(l+1)$ where  $j=1,...,l$, we have

\[
\left[S_{2n,l},\text{ }M_{2n,l}\right]\rightarrow_{d}\left[\frac{1}{l}\sum_{j=1}^{l}g(X_{\tau_{j}})\int K,~\mathbf{ MN}\Big(\mathbf{0},~\frac{\sigma_{u}^{2}}{l}\sum_{j=1}^{l}g(X_{\tau_{j}})g(X_{\tau_{j}})^{\prime}\int K^{2}\Big)\right].
\]
\end{rem}

\bigskip

Theorem \ref {lemma2}  provides limit theory for rescaled functionals
of nonstationary processes (i.e. $d_{n}^{-1}x_{k}$ as given \textbf{A3}).
For the purposes of regression analysis, limit theory for non rescaled
processes (i.e. $x_{k}$) is more relevant. Following Park and Phillips
(1999, 2001), we assume that the function $g(.)=[g_{1}(.),...,g_{p}(.)]^{\prime}$
is asymptotically homogeneous,
i.e. for large $\lambda$
\[
g_{i}(\lambda x)\approx\pi_{i}(\lambda)H_{i}(x),\text{ }i=1,...,p
\]
where $\pi_{i}$ (positive real valued function)\ is the ``asymptotic
order''\ of $g_{i}$\ and$\ H_{i}$ is the ``asymptotic homogeneous
function''\ of $g_{i}$ that is assumed continuous.
Several specifications of interest satisfy these conditions e.g. polynomial
functions, logarithmic, indicator functions and distribution type
of functions e.g. see Park and Phillips (2001) for more details. Set
$\pi\left(.\right):=$diag$\{\pi_{1}\left(.\right),...,\pi_{p}\left(.\right)\}$ and
$H(.)=[H_{1}(.),...,H_{p}(.)]^{\prime}$.
The following result is the counterpart of Theorem \ref {lemma2} for
additive transformations of non rescaled sequences.

\bigskip{}

\begin{thm}
\label{lemma3} Suppose that:
\begin{itemize}
\item[$(a)$] {\bf A1, A3} and {\bf A4} or {\bf A4$^*$} hold;
\item[$(b)$] for each $i=1,..,p$, there exists a continuous function
$H_{i}$ and $\pi_{i}:(0,\infty)\rightarrow(0,\infty)$, so that
\[
g_{i}(\lambda x)=\pi_{i}(\lambda)H_{i}(x)+R_{i}(\lambda,x),
\]
where $\left\vert R_{i}(\lambda,x)\right\vert \leq a_{i}(\lambda)(1+|x|^{\delta})$
for some $\delta>0$ and $a_{i}(\lambda)/\pi_{i}(\lambda)\rightarrow0,$
as $\lambda\rightarrow\infty$.

%\item[$(b)$] For each $i=1,..,p$ there exists a locally Riemann integrable function
%$H_{i}$ and $\pi_{i}:(0,\infty)\rightarrow(0,\infty)$, so that for
%$j=\left\{ 1,2\right\} $
%\[
%g_{i}(\lambda x)^{j}=\pi_{i}^{j}(\lambda)H_{i}^{j}(x)+R_{i,j}(\lambda,x),
%\]
%where $\left\vert R_{i,j}(\lambda,x)\right\vert \leq a_{i,j}(\lambda)(1+|x|^{\delta})$
%for some $\delta>0$ and $a_{i,j}(\lambda)/\pi_{i}^{j}(\lambda)\rightarrow0,$
%as $\lambda\rightarrow\infty$.
\end{itemize}
\noindent Then, as $n\rightarrow\infty$, we have
\begin{eqnarray}
&&\sum_{k=1}^{n}\pi\left(d_{n}\right)^{-1}g(x_{k})\Big\{ \sum_{j=1}^{l_{n}}K\left[c_{n}(k/n-\tau_{j})\right]\Big\} \Big[\frac{c_{n}}{nl_{n}},\text{ }\sqrt{\frac{c_{n}}{nl_{n}}}u_{k}\Big]\nonumber\\
&=&\sum_{k=1}^{n}H(X_{nk})\Big\{ \sum_{j=1}^{l_{n}}K\left[c_{n}(k/n-\tau_{j})\right]\Big\} \Big[\frac{c_{n}}{nl_{n}},\text{ }\sqrt{\frac{c_{n}}{nl_{n}}}u_{k}\Big]+o_{P}(1)\label{Lemma3a}\\
&\rightarrow_{d}&
\Big[\int_{0}^{1}H(X_{t})dt\int K,~ \mathbf{MN}\Big(\mathbf{0},~\sigma_{u}^{2}\int_{0}^{1}H(X_{t})H(X_{t})^{\prime}dt\int K^{2}\Big)\Big].\label{Lemma3b}
\end{eqnarray}
\end{thm}

\begin{rem} {\label{remark3a}}
 As noticed in Remark 3, if {\bf A4} (c) is replaced by  $\tau_{j}=j/(l+1)$ where  $j=1,...,l$,  we have
\begin{eqnarray*}
&&\sum_{k=1}^{n}\pi\left(d_{n}\right)^{-1}g(x_{k})\Big\{ \sum_{j=1}^{l}K\left[c_{n}(k/n-\tau_{j})\right]\Big\} \Big[\frac{c_{n}}{nl},\text{ }\sqrt{\frac{c_{n}}{nl}}u_{k}\Big]\nonumber\\
&\rightarrow_{d}&\left[\frac{1}{l}\sum_{j=1}^{l}H(X_{\tau_{j}})\int K,~\mathbf{ MN}\Big(\mathbf{0},~\frac{\sigma_{u}^{2}}{l}\sum_{j=1}^{l}H(X_{\tau_{j}})H(X_{\tau_{j}})^{\prime}\int K^{2}\Big)\right].
\end{eqnarray*}

\end{rem}

\begin{rem}
Suppose $K^{\ast}$ is a real function satisfying \textbf{A4} (a) or  \textbf{A4}$^*$ (a).  Let  $0<\tau^{\ast}<1$.
 Similar arguments as in the proof of Theorems \ref{lemma2} and \ref{lemma3}  show that, under the conditions of Theorem \ref{lemma3} with $g(.)=[g_1(.), g_2(.)]'$,
\be
\Big(\int_0^1H_1(X_{n, [nt]})dt, \ U_{1n},\ U_{2n}\Big) &\to_d & \Big(\int_0^1H_1(X_t)dt, \ \mathbf{MN}\left(\mathbf{0},\sigma_{u}^{2}\,V_1\right)\Big), \la {ad21} \\
\Big(\int_0^1H_1(X_{n, [nt]})dt,\  U_{1n},\ U_{3n}\Big) &\to_d & \Big(\int_0^1H_1(X_t)dt, \mathbf{MN}\left(\mathbf{0},\sigma_{u}^{2}\,V_2\right)\Big), \la {ad22}
\ee
 where
\bestar
U_{1n} &=&\sqrt{\frac{c_{n}}{n}}\sum_{k=1}^{n}\pi_2\left(d_{n}\right)^{-1}g_2(x_{k})\left\{ \frac{1}{\sqrt{l_n}}\sum_{j=1}^{l_n}K\left[c_{n}(k/n-\tau_{j})\right]\right\} u_{k},\\
U_{2n} &=&\sqrt{\frac{c_{n}}{n}}\sum_{k=1}^{n}\left\{ \frac{1}{\sqrt{l_n}}\sum_{j=1}^{l_n}K^*\left[c_{n}(k/n-\tau_{j})\right]\right\} u_{k}, \\
U_{3n} &=&\sqrt{\frac{c_{n}}{n}}\sum_{k=1}^{n}K^*\left[c_{n}(k/n-\tau^*)\right]\, u_{k},  \\
V_1 &=&  \begin{bmatrix}  %\left[\begin{array}{cc}
\int_{0}^{1}H_2^{2}\left(X_{t}\right)dt\int K^{2} & \int_{0}^{1}H_2\left(X_{t}\right)dt\, \int KK^*\\
\int_{0}^{1}H_2\left(X_{t}\right)dt\, \int KK^* & \int\left(K^{\ast}\right)^{2}
 \end{bmatrix}, \\  %\end{array}\right],\
&& \\
V_2 &=& \begin{bmatrix} %\left[\begin{array}{cc}
\int_{0}^{1}H_2^{2}\left(X_{t}\right)dt\int K^{2} & 0\\
0 & \int\left(K^{\ast}\right)^{2}
 \end{bmatrix}.% \end{array}\right].
\eestar
The limit results (\ref {ad21}) and (\ref {ad22}), together with Theorems \ref {lemma1} - \ref {lemma3}, will be utilised in Section 3 next.
\end{rem}

\section{LTLS estimation and inference}

The limit theory presented in the previous section is subsequently
utilised for deriving the properties of the LTLS estimator and a related
t-statistic. We consider nonlinear models of the form
\begin{equation}
y_{k}=\mu+\beta f(x_{k})+u_{k},\text{ }k=1,...,n,\label{model1}
\end{equation}
where $f$ is a known regression function $(\mu,\beta)$ unknown parameters
and the covariate $x_{k}$ can be nonstationary process or a stationary
one amenable to the limit theory of Theorem \ref {lemma1} or Theorem \ref {lemma2}
respectively. Further, $x_{k}$ is predetermined with respect to the
error $u_{t}$ in the sense $x_{k}$ is $\mathcal{F}_{k-1}$-measurable
and $\left\{ u_{k},\mathcal{F}_{k}\right\} $ is a martingale difference
(c.f. Assumptions {\bf A1-A3}). Similar nonlinear models with a predetermined
covariate have been considered for example by Park and Phillips (1999,
2001) and Chan and Wang (2015), in a parametric set up, and by Wang
and Phillips (2009a,b, 2011, 2012) in a nonparametric set-up.\footnote{Here we consider nonlinear models in $x_{k}$ only. Our results can
be generalised to models that are both nonlinear in $x_{k}$ and the
parameters along the lines of Chan and Wang (2015) for instance.}

Let $K$ be a kernel function  satisfying {\bf A4}(a) or {\bf A4$^*$}(a). Let $\tau_j=j/(l_n+1), j=1, ..., l_n$,   $c_{n}$ and $l_{n}$ be deterministic sequences satisfying {\bf A4}(b) and (c) or {\bf A4$^*$}(b) and (c). We also allow for $l_n$ to be a fixed constant.
Set
\begin{equation}
K_{kn}:=\sum_{j=1}^{l_{n}}K\left[c_{n}\left(k/n-\tau_{j}\right)\right]. \label{kernel1}
\end{equation}
Our aim is to estimate  the unknown parameter $\beta$ in (\ref{model1}) by
using the following instrument for $f(x_{k})$
\begin{eqnarray*}
Z_{kn}:=f_{k}K_{kn}:=f(x_{k})K_{kn}.
\end{eqnarray*}
As remarked in Section 1, due to the integrability of $K$, a trimming
effect applies around the chronological point(s) (\textit{cp(s)} hereafter)
$\tau_{j}$ which in turn reduces the signal of the OLS instrument
$f(x_{k})$. The reduction is more pronounced when the distance between
$k/n$ and$\ \tau_{j}$ is large, and/or the sequence $c_{n}$ diverges
fast. Clearly, for $K_{kn}=1$ we get the OLS estimator as a special
case. The reduction in the instrument signal enables an extended martingale given by Wang (2014) to operate. As a result the estimator under
consideration has a mixed Gaussian limit distribution, making pivotal inference possible.

A trimming method
is also crucial for demeaning $\left\{ y_{k}\right\} $ i.e. taking
into account the unknown intercept $\mu$. Let $K_{kn}^{\ast}$, $k=1,...,n$\ be
additive functionals of certain integrable kernel function.\
For any sequence $\left\{ a_{k}\right\} _{k=1}^{n}$\ let
\begin{equation}
\overline{a}:=\frac{\sum_{k=1}^{n}a_{k}K_{kn}^{\ast}}{\sum_{k=1}^{n}K_{kn}^{\ast}}\text{\qquad and}\qquad\overline{a}_{k}:=a_{k}-\overline{a}\text{.}\label{demeaning}
\end{equation}
\ We will consider two possibilities for $K_{kn}^{\ast}$. Either
\begin{equation}
K_{kn}^{\ast}:=\sum_{j=1}^{l_{n}}K^{\ast}\left[c_{n}\left(k/n-\tau_{j}\right)\right]\text{\qquad\ or\qquad\ }K_{kn}^{\ast}:=K^{\ast}\left[c_{n}\left(k/n-\tau^{\ast}\right)\right],\label{kernel2}
\end{equation}
where $K^{\ast}$ satisfies {\bf A4}(a), $\tau_j=j/(l_n+1), j=1,2,..., l_n,$ are given above and  $\ 0<\tau^{\ast}<1$. The first term in (\ref{demeaning}) involves
a trimmed sample mean around an array of several \textit{cps}, whilst
the second is a trimmed sample mean based on a single fixed \textit{cp}.\ Define
the LTLS estimator as
\[
\hat{\beta}:=\frac{\sum_{k=1}^{n}Z_{kn}\overline{y}_{k}}{\sum_{k=1}^{n}Z_{kn}\overline{f}_{k}}.
\]
The employment of a ``trimmed''\ sample mean is crucial for obtaining
mixed Gaussian limit theory. Notice that
\[
\hat{\beta}=\beta+\frac{1}{\sum_{k=1}^{n}Z_{kn}\overline{f}_{k}}\left\{ \sum_{k=1}^{n}f_{k}K_{kn}u_{k}-\frac{\left(\sum_{k=1}^{n}f_{k}K_{kn}\right)\sum_{k=1}^{n}K_{kn}^{\ast}u_{k}}{\sum_{k=1}^{n}K_{kn}^{\ast}}\right\} .
\]
For nonstationary $x_{k}$ the two martingale terms shown above converge
jointly to a bivariate mixed Gaussian limit. In particular,
\[
\left[\sqrt{\frac{c_{n}}{n}}\sum_{k=1}^{n}f\left(d_{n}^{-1}x_{k}\right)K_{kn}u_{k},\ \sqrt{\frac{c_{n}}{n}}\sum_{k=1}^{n}K_{kn}^{\ast}u_{k}\right]\rightarrow_{d} \mathbf{MN}\left(\mathbf{0},V\right),
\]
for some random matrix $V$. Note that if instead the standard demeaning
was employed (i.e. $K^{\ast}=1$), then
\[
\left[\sqrt{\frac{c_{n}}{n}}\sum_{k=1}^{n}f\left(d_{n}^{-1}x_{k}\right)K_{kn}u_{k},\ \frac{1}{\sqrt{n}}\sum_{k=1}^{n}u_{k}\right]\nrightarrow_{d}\mathbf{MN}\left(\mathbf{0},V\right),
\]
for some random matrix $V$, despite the fact that each of the components
on the l.h.s. above converges weakly to some (mixed) Gaussian limit.

To investigate  the limit properties of the LTLS estimator $\hat{\beta}$ in detail, set
\[
\lambda_{n}:=\frac{nl_{n}}{c_{n}}\text{\qquad\ and \qquad}\lambda_{n}^{\ast}:=\frac{nl_{n}^{\ast}}{c_{n}},
\]
where
\[
l_{n}^{\ast}:=\left\{ \begin{array}{l}
l_{n}\text{, if }K_{kn}^{\ast}=\sum_{j=1}^{l_{n}}K^{\ast}\left[c_{n}\left(k/n-\tau_{j}\right)\right]\\
1\text{, if }K_{kn}^{\ast}=K^{\ast}\left[c_{n}\left(k/n-\tau^{\ast}\right)\right]
\end{array}\right..
\]
The sequences $\lambda_{n}$, $\lambda_{n}^{\ast}$ give the order
of the terms\footnote{Note that by standard arguments (Euler summation)
\[
\sum_{k=1}^{n}K_{kn}^{\ast}\sim\frac{nl_{n}^{\ast}}{c_{n}}\int K^{\ast}.
\]
} $\sum_{k=1}^{n}K_{kn}$ and $\sum_{k=1}^{n}K_{kn}^{\ast}$ which
in turn\
determine the convergence rate of the LTLS estimator.
Further set
\begin{quote}
$R^{\ast}=1$ and $Q^{\ast}=\int KK^{\ast}$ if $l_{n}^{\ast}=l_{n}$;\ \
$R^{\ast}=Q^{\ast}=0$ if $l_{n}^{\ast}=1$.
\end{quote}

We have the following main results for the asymptotics of the LTLS estimator $\hat{\beta}$. Theorem \ref {Thm1} is for stationary regressor. Limit theory in nonstationary case is given in Theorem~\ref {Thm2}.

\begin{thm}
\label{Thm1} Suppose that:
\begin{itemize}
\item[$(a)$]  \textbf{A1}, \textbf{A2}  with $g=f$,  and \textbf{A4} or \textbf{A4$^*$} hold;
\item[$(b)$] $K^{\ast}$ satisfies \textbf{A4}(a) or \textbf{A4$^*$}(a)   and  $0<\tau^*<1$.
\end{itemize}
Then, as  $n\rightarrow\infty$, we have
\begin{eqnarray}
\sqrt{\lambda_{n}^{\ast}}\left(\hat{\beta}-\beta\right)&\rightarrow_{d}&
\sigma_{u}\,   \mathbf{N}\left({0}, \,\Omega^{-2}\,LML'\right),
\label{thm1}
\end{eqnarray}
where $\Omega=\left\{ Ef^2(x_{1})-\left[Ef(x_{1})\right]^{2}\right\} \int K$, $L=\left(R^{\ast},~-Ef(x_{1})\int K/\int K^{\ast}\right) $ and
\bestar
M = \left[\begin{array}{cc}
Ef(x_{1})^{2}\int K^{2} & Ef(x_{1})Q^{\ast}\\
Ef(x_{1})Q^{\ast} & \int\left(K^{\ast}\right)^{2}
\end{array}\right].
\eestar
\end{thm}

\begin{thm}
\label{Thm2} Suppose that
\begin{itemize}
\item[$(a)$]  \textbf{A1}, \textbf{A3} and \textbf{A4} or \textbf{A4$^*$} hold;
\item[$(b)$] $f(x)$ is an asymptotically homogeneous function, i.e., there exists
a continuous function $H$ and $\pi:\left(0,\infty\right)\rightarrow\left(0,\infty\right)$
such that
\[
f(\lambda x)=\pi(\lambda)H(x)+R(\lambda,x),
\]
where $\left\vert R(\lambda,x)\right\vert \leq a(\lambda)(1+|x|^{\delta})$
for some $\delta>0$ and $a(\lambda)/\pi(\lambda)\rightarrow0,$
as $\lambda\rightarrow\infty$;
\item[$(c)$] $K^{\ast}$ satisfies \textbf{A4}(a) or \textbf{A4$^*$}(a)   and  $0<\tau^*<1$.
\end{itemize}
Then, as $n\rightarrow\infty$,
\begin{eqnarray}
\sqrt{\lambda_{n}^{\ast}}\pi(d_{n})\left(\hat{\beta}-\beta\right)&\rightarrow_{d}& \sigma_{u}\, \mathbf{MN}\left({0},~\Big(C\int K\Big)^{-2}\,AVA'\right),~~~~\label{thm2}
\end{eqnarray}
where
\bestar
C&=& \left\{ \begin{array}{ll}
 \int_{0}^{1}H^{2}(X_{t})dt-\left[\int_{0}^{1}H(X_{t})dt\right]^{2}, &\text{ if }K_{kn}^{\ast}=\sum_{j=1}^{l_{n}}K^{\ast}\left[c_{n}\left(k/n-\tau_{j}\right)\right],\\
 \int_{0}^{1}H^{2}(X_{t})dt-\big[\int_{0}^{1}H(X_{t})dt\big]\, H(X_{\tau^*}), &\text{ if }K_{kn}^{\ast}=K^{\ast}\left[c_{n}\left(k/n-\tau^{\ast}\right)\right],
\end{array}\right. \\
A &=& \left[R^{\ast},~-\int_{0}^{1}H(X_{t})dt\int K/\int K^{\ast}\right], \quad \mbox{and} \\
V &=&  \left[\begin{array}{ll}
\int_{0}^{1}H^{2}(X_{t})dt\int K^{2} & \int_{0}^{1}H(X_{t})dt\,Q^{\ast}\\
\int_{0}^{1}H(X_{t})dt\, Q^{\ast} & \int\left(K^{\ast}\right)^{2}
\end{array}\right].
\eestar

\end{thm}
\bigskip{}

\begin{rem}
\label{remark4}
Due to the fact that $\sqrt{\lambda_{n}^{\ast}}=o\left(\sqrt{n}\right)$, the convergence rate in LTLS for both  stationary and nonstationary regressor is slower in comparison with that of the OLS estimator.
\end{rem}

 \begin{rem} When a single \textit{cp} is used in demeaning $y_{k}$, we have $R^{\ast},Q^{\ast}=0$. In this case, the right hand side of (\ref {thm2}) becomes
\[
\frac{-\int_{0}^{1}H(X_{t})dt/\int K^{\ast}}{ \int_{0}^{1}H^{2}(X_{t})dt-\left[\int_{0}^{1}H(X_{t})dt\right]\, H(X_{\tau^*})}\times \mathbf{N}\left(0,~\sigma_{u}^{2}\int\left(K^{\ast}\right)^{2}\right).
\]
Simulations presented in Section 4 show that, in finite samples, superior
performance is obtained for certain configuration that involves multiple
\textit{cps} for the instrumentation of $x_{k}$\ (i.e. $K$)\ and
a single \textit{cp} for demeaning $y_{k}$ (i.e. $K^{\ast}$).\ An
analogous result can be established when the opposite holds i.e. a
single \textit{cp} ($\tau$, say) is used for the instrumentation of $x_{k}$ (i.e.
$K$)\ and multiple \textit{cps} (i.e. $K^*$) are\ used for
demeaning. In particular, in the latter case it can be shown that
the limit distribution (nonstationary $x_{k}$) is
\[
\frac{-H(X_{\tau})/ \int K^{\ast}}{H^2(X_{\tau})-\left[\int_{0}^{1}H(X_{t})dt\right] \, H(X_{\tau})}\times \mathbf{N}\left(0,\sigma_{u}^{2}\int\left(K^{\ast}\right)^{2}\right).
\]
We do not consider this possibility in the theorems shown above explicitly
in order avoid more complex exposition.

%Finally, note that in all cases the convergence rate is $\sqrt{\lambda_{n}^{\ast}}\pi(d_{n})$.
\end{rem}
\bigskip{}

To end this section, we consider the following $t$-statistic for the hypothesis
$H_{0}:\beta=\beta_{0}$ (for some $\beta_{0}\in\mathbb{R}$)

\begin{equation}
\hat{T}:=\mathcal{C}_{n}\frac{\hat{\beta}-\beta_{0}}{\sqrt{\tilde{\sigma}^{2}\mathcal{A}_{n}\mathcal{V}_{n}\mathcal{A}_{n}^{\prime}}},\label{t-statistic}
\end{equation}
where
\[
\mathcal{A}_{n}:=\left[1,\text{ }-\frac{\sum_{k=1}^{n}f_kK_{kn}}{\sum_{k=1}^{n}K_{kn}^{\ast}}\right],~~\text{ }\mathcal{C}_{n}:=\sum_{k=1}^{n}Z_{kn}\overline{f}_{k},
\]
\[
\mathcal{V}_{n}:=\left[\begin{array}{ll}
\sum_{k=1}^{n}K_{kn}^{2}f_{k}^{2} & \sum_{k=1}^{n}K_{kn}^{\ast}K_{kn}f_{k}\\
\sum_{k=1}^{n}K_{kn}^{\ast}K_{kn}f_{k} & \sum_{k=1}^{n}\left(K_{kn}^{\ast}\right)^{2}
\end{array}\right],
\]
and $\tilde{\sigma}^{2}:=n^{-1}\sum_{k=1}^{n}\tilde{u}_{k}^{2}$,
where $\tilde{u}_{k}$ are residuals from OLS estimation of (\ref{model1}).
The limit properties of $\hat{T}$ under the null hypothesis are demonstrated
by Theorem \ref{thm3} below. \bigskip{}

\begin{thm}
\label{thm3}Suppose that either conditions of Theorem \ref{Thm1}
or Theorem \ref{Thm2} hold. Then under $H_{0}:\beta=\beta_{0}$,
\[
\hat{T}\rightarrow_{d}N(0,1).
\]
\end{thm}

\begin{rem}  Note that the limit distribution of the test statistic under the null hypothesis is standard normal for both stationary and nonstationary regressors. Under the alternative hypothesis, the divergence rate of $\hat{T}$ is determined by the convergence rate of the LTLS estimator. In particular, for stationary $x_k$ it can be easily seen that $\hat{T}=O_P(\sqrt{\lambda_{n}^{\ast}})$. On the other hand in the nonstationary case we have $\hat{T}=O_P(\sqrt{\lambda_{n}^{\ast}}\pi(d_{n}))$, where $d_{n}=\sqrt{n}$ for $x_k$ NI or I(1) and $d_{n}=n^{d}$, $1/2<d<3/2$. Therefore, faster divergence rate is attained for more persistence processes. This fact is also corroborated by our simulation results (see Figures 1-3).
\end{rem}

%appendix

\section{Simulations}

We next investigate the final sample performance of the t-test based
on the LTLS estimator. In particular we test the hypothesis $H_{0}:\beta=0$
against $H_{1}:\beta\neq0$ at 5\% significance level. The vector
$\left[\xi_{k},u_{k}\right]$ process is generated by
\[
\left[\begin{array}{c}
\xi_{k}\\
u_{k}
\end{array}\right]\sim i.d.N\left(\mathbf{0},\left[\begin{array}{cc}
1 & \delta\\
\delta & 1
\end{array}\right]\right)
\]
Further, for $k=1,...,n$ the process$\ \left\{ y_{k}\right\} $ is
generated by
\[
y_{k+1}=\beta x_{k}+u_{k+1}
\]
where $\left\{ x_{k}\right\} $ is either a NI array of the form
\begin{equation}
x_{k}=\left(1+\frac{c}{n}\right)x_{k-1}+\xi_{k},\label{sims1}
\end{equation}
with $c\leq0$ and $x_{0}=0$\ or a type II fractional process (e.g.
see Robinson and Hualde, 2003) of the form
\begin{equation}
\left(I-L\right)^{d}x_{k}=\xi_{k}1\left\{ k\geq1\right\} .\label{sims2}
\end{equation}

Let $\varphi_{\varsigma^{2}}(x)$ be the density of a $N\left(0,\varsigma^{2}\right)$
variate. Next, set $\tilde{\sigma}_{u}^{2}=n^{-1}\sum_{k=1}^{n}\tilde{u}_{k}^{2}$,
$\tilde{\sigma}_{\xi}^{2}=n^{-1}\sum_{k=1}^{n}\tilde{\xi}_{k}^{2}$,$\ \tilde{\delta}=\frac{n^{-1}\sum_{k=1}^{n}\tilde{u}_{k}\tilde{\xi}_{k}}{\sqrt{\tilde{\sigma}_{u}^{2}\tilde{\sigma}_{\xi}^{2}}}$,
where $\tilde{u}_{t}$ and $\tilde{\xi}_{k}$ are OLS residuals from
the regressions
\[
y_{k+1}=\tilde{\mu}+\tilde{\beta}x_{k}+\tilde{u}_{k+1}\qquad\text{and}\qquad x_{k}=\tilde{\mu}_{x}+\tilde{\rho}x_{k-1}+\tilde{\xi}_{k}
\]
respectively.\ Finally, $\left\{ \tau_{j}\right\} _{j=1}^{l_{n}}$
are equispaced points on $(0,1)$.

We consider 3 set-ups for kernel functionals and \textit{cps}.

\bigskip{}

\begin{itemize}
\item[S1] (set-up 1) $K_{kn}=\sum_{j=1}^{l_{n}}K\left[c_{n}\left(k/n-\tau_{j}\right)\right]$,
$K_{kn}^{\ast}=\sum_{j=1}^{l_{n}}K^{\ast}\left[c_{n}\left(k/n-\tau_{j}\right)\right]$,
$K(x)=\varphi_{0.1}(x)^{1/2}$, $K(x)^{\ast}=\varphi_{1}(x)^{1/2}$,
$c_{n}=n^{0.95}$, $l_{n}=c_{n}^{0.7}$.
\item[S2] (set-up 2) $K_{kn}=\sum_{j=1}^{l_{n}}K\left[c_{n}\left(k/n-\tau_{j}\right)\right]$,
$K_{kn}^{\ast}=\sum_{j=1}^{l_{n}}K^{\ast}\left[c_{n}\left(k/n-\tau_{j}\right)\right]$,
$K(x)=\varphi_{0.1}(x)^{1/2}$, $K(x)^{\ast}=\varphi_{1}(x)^{1/2}$,
$c_{n}=n^{0.95}$, $l_{n}=c_{n}^{\hat{\alpha}}$, $\hat{\alpha}=1-0.45\left\vert \tilde{\delta}\right\vert $.
\item[S3] (set-up 3) $K_{kn}=\sum_{j=1}^{l_{n}}K\left[c_{n}\left(k/n-\tau_{j}\right)\right]$,
$K_{kn}^{\ast}=K^{\ast}\left[c_{n}\left(k/n-0.5\right)\right]$, $K(x)=\varphi_{\hat{\varsigma}^{2}}(x)$,
$K(x)^{\ast}=\varphi_{\hat{\varsigma}^{2}}(x)^{1/2}$, $\hat{\varsigma}^{2}=\tilde{\sigma}_{u}^{2}\left(0.1+0.9\left\vert \tilde{\delta}\right\vert \right)$,
$c_{n}=n^{\hat{\alpha}}$, $\hat{\alpha}=-0.1+0.15\left\vert \tilde{\delta}\right\vert $,
$l_{n}=\log n$.
\end{itemize}
\bigskip{}

In S1 and S2 multiple \textit{cps} are used for both $K_{kn}$\ and
$K_{kn}^{\ast}$\ whilst in S3 $K_{kn}^{\ast}$\ involves a single
\textit{cp}. Contrary to S1, in S2 a data driven approach is followed
for the determination of the number of \textit{cps} ($l_{n}$). As
remarked in Section 1, a small $c_{n}$ and/or large number of \textit{cps}\ results
in a LTLS estimator approximately equal to the OLS estimator. The
OLS estimator in general has a good power properties but is severely
oversized when endogeneity is strong (i.e. when $\left\vert \delta\right\vert $\ is
close to one). In S2 a large number of \textit{cps} is utilised when
endogeneity is weak whilst for $l_{n}$ drops as $\left\vert \delta\right\vert $
approaches one. A similar data-driven approach is utilised in S3.
In this case $c_{n}$ is very small (vanishing) for $\delta$ close
to zero, whilst $c_{n}$ is large (diverging) for $\left\vert \delta\right\vert $
close to one. Further, in S3 the choice of the kernel variance is also
data driven. Preliminary simulations have shown that superior performance
is attained when $\varsigma^{2}=0.1$ for $\delta\approx0$ and $\varsigma^{2}=1$
for $\left\vert \delta\right\vert \approx1$. Therefore, $\hat{\varsigma}^{2}=\tilde{\sigma}_{u}^{2}\left(0.1+0.9\left\vert \tilde{\delta}\right\vert \right)$
provides an interpolation between these values based on the actual
data.

For S1 and S2 we use the test statistic of (\ref{t-statistic}). For
S3 we use $\mathcal{A}_{n}^{\ast}:=\left[1,\text{ }-\frac{\sum_{k=1}^{n}f(x_{k})K_{kn}^{\ast}}{\sum_{k=1}^{n}K_{kn}^{\ast}}\right]$
instead of $\mathcal{A}_{n}$ in (\ref{t-statistic}). Note that given
the configuration of S3, in the nonstationary case, $\frac{\sum_{k=1}^{n}f(x_{k})K_{kn}^{\ast}}{\sum_{k=1}^{n}K_{kn}^{\ast}}=O_{p}\left(\pi_{f}(d_{n})\right)$
whilst $\frac{\sum_{k=1}^{n}f(x_{k})K_{kn}}{\sum_{k=1}^{n}K_{kn}^{\ast}}$
that appears in $\mathcal{A}_{n}\ $is $O_{p}\left(\pi_{f}(d_{n})\log n\right)$.
Therefore, the employment of $\mathcal{A}_{n}^{\ast}$ results in
giving less weight in the term that correspondents to the studentisation
of the intercept correction. Note that in infinite samples the utilisation
of $\mathcal{A}_{n}^{\ast}$ does not result in a consistent estimator
for the limit variance of $\hat{\beta}-\beta$. Nevertheless, our
simulation results reveal that in finite samples a superior performance
in attained when $\mathcal{A}_{n}^{\ast}$ is employed.

Table 1 shows the size properties of the LTLS based t-tests, for the
case the regressor is a NI array generated by (\ref{sims1}). The
number of replication paths is set 10,000 throughout. We also consider
the IVX based test (see eq. (20) in Kostakis et al, 2015) and the
OLS based t-test. We allow for several values of the correlation parameter
($\delta=\{-.95,-.5,0,.5,.95\}$) the near to unity parameter ($c=\{0,-5,-10,-20,-50\}$)
and sample size ($n=\{250,500,750,1000\}$). We use the notation T1,
T2 and T3 to denote the LTLS t-statistics that correspond to set-ups
S1, S2 and S3 respectively. In general, all LTLS based test exhibit
good size control. Under S1 and S2 the tests are moderately oversized
for small samples sizes when $c=0$ and correlation $\left\vert \delta\right\vert =.95$.
Figure 1 and Figure 2 show the empirical power ($n=250$) of the LTLS
and IVX tests for $c=0$ and $c=-20$ respectively. It can be seen
from these figures that T3 attains better performance than the other
LTLS based tests under consideration (i.e. T1 and T2). In particular,
the performance of the the LTLS t-test under S3 is almost identical
to that of the IVX based test. This is somewhat surprising given that
under S3 the studentisation used does not lead to a consistent estimator
for the limit variance of the LTLS estimator. As noted above, under
S3 the term that provides studentisation to the intercept correction
is of slightly smaller order of magnitude (i.e. $\log n$) than the
corresponding term in $\mathcal{A}_{n}$. The simulation study provided
suggests that this misbalancing leads to some finite sample improvement.
Hosseinkouchack and Demetrescu (2019) provide finite sample improvements
to the the IVX method. These authors show that the IVX t-statistic
distribution is skewed relative to the $N(0,1)$ in finite samples
when endogeneity is strong. It is reasonable to expect that a similar
phenomenon holds for the LTLS distribution in finite samples. It seems
that the utilisation $\mathcal{A}_{n}^{\ast}$ provides a rebalancing
to the test statistic that corrects for deviations from the standard
normal distribution. A rigorous analysis for the performance of the
T3 in finite samples, would require developing higher order limit
theory. A development in this direction is challenging from a technical
point of view and will be left for future work.

We next consider the case where the regressor is a non stationary
fractional process (i.e. (\ref{sims2})). The finite sample size performance
of T3 and the LS based test procedure are shown in Table 2.\footnote{Preliminary simulation results show that the performance of T1 and
T2 in the fractional case is comparable to that in the NI case.} It can be seen from Table 2 that the T3 test provides good size control
for a wide range levels in persistence and endogeneity. On the other
hand LS based test may exhibit serious oversizing. In particular,
for $\delta=-.95$ empirical size ranges from three times ($d=0.75$)
to six times ($d=1.2$) the nominal one. Finally, Figure 3 shows the
finite power of T3 for $n=250$, $d=\{0.8,1,1.2\}$ and $\delta=\{0,-.5,-.95\}$.
As expected, better power performance is attained for more persistent
regressors.

\clearpage\thispagestyle{empty} \begin{landscape}
\begin{table}[htbp]
\centering \caption{Empirical Size (NI regressor; 5\% Nominal Size)}
\resizebox{1.6\textwidth}{!}{

\begin{tabular}{l||rrrrr|rrrrr|rrrrr|rrrrr|rrrrr}
\multicolumn{1}{l}{} & \multicolumn{5}{c}{} & \multicolumn{5}{c}{} & \multicolumn{5}{c}{} & \multicolumn{5}{c}{} & \multicolumn{5}{c}{}\tabularnewline
\hline
\hline
\multicolumn{1}{l||}{$c=0$} & \multicolumn{5}{c|}{$\delta=-0.95$} & \multicolumn{5}{c|}{$\delta=-0.5$} & \multicolumn{5}{c|}{$\delta=0$} & \multicolumn{5}{c|}{$\delta=0.5$} & \multicolumn{5}{c}{$\delta=0.95$}\tabularnewline
\multicolumn{1}{l||}{\textit{n}} & \multicolumn{1}{l}{T1} & \multicolumn{1}{l}{T2} & \multicolumn{1}{l}{T3} & \multicolumn{1}{l}{IVX} & \multicolumn{1}{l|}{OLS} & \multicolumn{1}{l}{T1} & \multicolumn{1}{l}{T2} & \multicolumn{1}{l}{T3} & \multicolumn{1}{l}{IVX} & \multicolumn{1}{l|}{OLS} & \multicolumn{1}{l}{T1} & \multicolumn{1}{l}{T2} & \multicolumn{1}{l}{T3} & \multicolumn{1}{l}{IVX} & \multicolumn{1}{l|}{OLS} & \multicolumn{1}{l}{T1} & \multicolumn{1}{l}{T2} & \multicolumn{1}{l}{T3} & \multicolumn{1}{l}{IVX} & \multicolumn{1}{l|}{OLS} & \multicolumn{1}{l}{T1} & \multicolumn{1}{l}{T2} & \multicolumn{1}{l}{T3} & \multicolumn{1}{l}{IVX} & \multicolumn{1}{l}{OLS}\tabularnewline
250 & 0.084 & 0.095  & 0.060  & 0.059  & 0.278  & 0.059  & 0.074  & 0.057  & 0.056  & 0.117  & 0.051  & 0.052  & 0.045 & 0.050 & 0.053 & 0.061 & 0.075 & 0.052 & 0.056 & 0.113 & 0.087 & 0.096 & 0.064 & 0.061 & 0.295\tabularnewline
500 & 0.077 & 0.078 & 0.062 & 0.062 & 0.287 & 0.059 & 0.067 & 0.051 & 0.054 & 0.114 & 0.054 & 0.053 & 0.046 & 0.054 & 0.054 & 0.060 & 0.076 & 0.057 & 0.058 & 0.116 & 0.080 & 0.083 & 0.058 & 0.055 & 0.279\tabularnewline
750 & 0.076 & 0.069 & 0.062 & 0.058  & 0.272  & 0.059  & 0.065  & 0.051  & 0.052  & 0.109  & 0.052  & 0.050  & 0.042  & 0.050  & 0.051  & 0.059  & 0.062  & 0.054  & 0.055  & 0.111  & 0.080  & 0.068  & 0.063  & 0.057 & 0.277\tabularnewline
1000 & 0.070 & 0.067 & 0.059 & 0.053 & 0.278 & 0.054 & 0.064 & 0.051 & 0.051 & 0.111 & 0.049 & 0.048 & 0.046 & 0.051 & 0.053 & 0.059 & 0.067 & 0.052 & 0.050 & 0.108 & 0.075 & 0.062 & 0.058 & 0.053 & 0.277\tabularnewline
\hline
\multicolumn{1}{l||}{$c=-5$} & \multicolumn{5}{c|}{$\delta=-0.95$} & \multicolumn{5}{c|}{$\delta=-0.5$} & \multicolumn{5}{c|}{$\delta=0$} & \multicolumn{5}{c|}{$\delta=0.5$} & \multicolumn{5}{c}{$\delta=0.95$}\tabularnewline
\multicolumn{1}{l||}{\textit{n}} & \multicolumn{1}{l}{T1} & \multicolumn{1}{l}{T2} & \multicolumn{1}{l}{T3} & \multicolumn{1}{l}{IVX} & \multicolumn{1}{l|}{OLS} & \multicolumn{1}{l}{T1} & \multicolumn{1}{l}{T2} & \multicolumn{1}{l}{T3} & \multicolumn{1}{l}{IVX} & \multicolumn{1}{l|}{OLS} & \multicolumn{1}{l}{T1} & \multicolumn{1}{l}{T2} & \multicolumn{1}{l}{T3} & \multicolumn{1}{l}{IVX} & \multicolumn{1}{l|}{OLS} & \multicolumn{1}{l}{T1} & \multicolumn{1}{l}{T2} & \multicolumn{1}{l}{T3} & \multicolumn{1}{l}{IVX} & \multicolumn{1}{l|}{OLS} & \multicolumn{1}{l}{T1} & \multicolumn{1}{l}{T2} & \multicolumn{1}{l}{T3} & \multicolumn{1}{l}{IVX} & \multicolumn{1}{l}{OLS}\tabularnewline
250  & 0.061  & 0.069  & 0.060  & 0.062  & 0.116  & 0.051  & 0.061  & 0.056  & 0.056  & 0.072  & 0.050  & 0.052  & 0.051  & 0.050  & 0.051  & 0.057  & 0.065  & 0.056  & 0.059  & 0.074  & 0.068  & 0.070  & 0.066  & 0.066  & 0.123 \tabularnewline
500  & 0.060  & 0.067  & 0.062  & 0.063  & 0.117  & 0.051  & 0.060  & 0.056  & 0.059  & 0.073  & 0.051  & 0.055  & 0.051  & 0.052  & 0.054  & 0.056  & 0.061  & 0.057  & 0.057  & 0.071  & 0.062  & 0.060  & 0.058  & 0.058  & 0.116 \tabularnewline
750  & 0.063  & 0.058  & 0.062  & 0.060  & 0.116  & 0.058  & 0.056  & 0.056  & 0.059  & 0.070  & 0.056  & 0.055  & 0.052  & 0.056  & 0.053  & 0.059  & 0.055  & 0.059  & 0.058  & 0.073  & 0.065  & 0.063  & 0.062  & 0.062  & 0.119 \tabularnewline
1000  & 0.058  & 0.055  & 0.059  & 0.060  & 0.116  & 0.049  & 0.059  & 0.052  & 0.054  & 0.066  & 0.047  & 0.052  & 0.050  & 0.050  & 0.051  & 0.050  & 0.056  & 0.053  & 0.052  & 0.066  & 0.059  & 0.056  & 0.057  & 0.058  & 0.115 \tabularnewline
\hline
\multicolumn{1}{l||}{$c=-10$} & \multicolumn{5}{c|}{$\delta=-0.95$} & \multicolumn{5}{c|}{$\delta=-0.5$} & \multicolumn{5}{c|}{$\delta=0$} & \multicolumn{5}{c|}{$\delta=0.5$} & \multicolumn{5}{c}{$\delta=0.95$}\tabularnewline
\multicolumn{1}{l||}{\textit{n}} & \multicolumn{1}{l}{T1} & \multicolumn{1}{l}{T2} & \multicolumn{1}{l}{T3} & \multicolumn{1}{l}{IVX} & \multicolumn{1}{l|}{OLS} & \multicolumn{1}{l}{T1} & \multicolumn{1}{l}{T2} & \multicolumn{1}{l}{T3} & \multicolumn{1}{l}{IVX} & \multicolumn{1}{l|}{OLS} & \multicolumn{1}{l}{T1} & \multicolumn{1}{l}{T2} & \multicolumn{1}{l}{T3} & \multicolumn{1}{l}{IVX} & \multicolumn{1}{l|}{OLS} & \multicolumn{1}{l}{T1} & \multicolumn{1}{l}{T2} & \multicolumn{1}{l}{T3} & \multicolumn{1}{l}{IVX} & \multicolumn{1}{l|}{OLS} & \multicolumn{1}{l}{T1} & \multicolumn{1}{l}{T2} & \multicolumn{1}{l}{T3} & \multicolumn{1}{l}{IVX} & \multicolumn{1}{l}{OLS}\tabularnewline
250  & 0.058  & 0.067  & 0.058  & 0.062  & 0.086  & 0.051  & 0.058  & 0.054  & 0.055  & 0.063  & 0.049  & 0.050  & 0.050  & 0.051  & 0.052  & 0.056  & 0.061  & 0.057  & 0.057  & 0.063  & 0.063  & 0.059  & 0.066  & 0.065  & 0.090 \tabularnewline
500  & 0.058  & 0.053  & 0.061  & 0.063  & 0.088  & 0.051  & 0.060  & 0.058  & 0.058  & 0.065  & 0.047  & 0.056  & 0.051  & 0.052  & 0.052  & 0.050  & 0.061  & 0.054  & 0.055  & 0.060  & 0.056  & 0.056  & 0.059  & 0.057  & 0.085 \tabularnewline
750  & 0.058  & 0.054  & 0.061  & 0.060  & 0.087  & 0.058  & 0.053  & 0.058  & 0.056  & 0.064  & 0.055  & 0.054  & 0.053  & 0.056  & 0.053  & 0.056  & 0.059  & 0.057  & 0.055  & 0.062  & 0.058  & 0.057  & 0.062  & 0.062  & 0.088 \tabularnewline
1000  & 0.053  & 0.052  & 0.058  & 0.058  & 0.084  & 0.049  & 0.052  & 0.053  & 0.053  & 0.059  & 0.046  & 0.048  & 0.048  & 0.050  & 0.051  & 0.049  & 0.056  & 0.050  & 0.051  & 0.058  & 0.054  & 0.055  & 0.058  & 0.058  & 0.088 \tabularnewline
\hline
\multicolumn{1}{l||}{$c=-20$} & \multicolumn{5}{c|}{$\delta=-0.95$} & \multicolumn{5}{c|}{$\delta=-0.5$} & \multicolumn{5}{c|}{$\delta=0$} & \multicolumn{5}{c|}{$\delta$$=0.5$} & \multicolumn{5}{c}{$\delta=0.95$}\tabularnewline
\multicolumn{1}{l||}{\textit{n}} & \multicolumn{1}{l}{T1} & \multicolumn{1}{l}{T2} & \multicolumn{1}{l}{T3} & \multicolumn{1}{l}{IVX} & \multicolumn{1}{l|}{OLS} & \multicolumn{1}{l}{T1} & \multicolumn{1}{l}{T2} & \multicolumn{1}{l}{T3} & \multicolumn{1}{l}{IVX} & \multicolumn{1}{l|}{OLS} & \multicolumn{1}{l}{T1} & \multicolumn{1}{l}{T2} & \multicolumn{1}{l}{T3} & \multicolumn{1}{l}{IVX} & \multicolumn{1}{l|}{OLS} & \multicolumn{1}{l}{T1} & \multicolumn{1}{l}{T2} & \multicolumn{1}{l}{T3} & \multicolumn{1}{l}{IVX} & \multicolumn{1}{l|}{OLS} & \multicolumn{1}{l}{T1} & \multicolumn{1}{l}{T2} & \multicolumn{1}{l}{T3} & \multicolumn{1}{l}{IVX} & \multicolumn{1}{l}{OLS}\tabularnewline
250  & 0.056  & 0.052  & 0.057  & 0.060  & 0.069  & 0.052  & 0.054  & 0.052  & 0.051  & 0.057  & 0.051  & 0.049  & 0.051  & 0.050  & 0.050  & 0.055  & 0.057  & 0.055  & 0.055  & 0.058  & 0.061  & 0.056  & 0.059  & 0.060  & 0.071 \tabularnewline
500  & 0.054  & 0.055  & 0.058  & 0.060  & 0.072  & 0.050  & 0.055  & 0.055  & 0.054  & 0.058  & 0.048  & 0.056  & 0.051  & 0.051  & 0.052  & 0.049  & 0.053  & 0.053  & 0.055  & 0.058  & 0.053  & 0.052  & 0.054  & 0.056  & 0.067 \tabularnewline
750  & 0.053  & 0.053  & 0.060  & 0.059  & 0.071  & 0.056  & 0.055  & 0.057  & 0.060  & 0.060  & 0.052  & 0.054  & 0.055  & 0.056  & 0.053  & 0.056  & 0.055  & 0.055  & 0.055  & 0.058  & 0.057  & 0.054  & 0.061  & 0.062  & 0.074 \tabularnewline
1000  & 0.052  & 0.052  & 0.057  & 0.057  & 0.071  & 0.047  & 0.057  & 0.052  & 0.050  & 0.056  & 0.048  & 0.049  & 0.048  & 0.048  & 0.049  & 0.048  & 0.050  & 0.048  & 0.049  & 0.053  & 0.052  & 0.055  & 0.057  & 0.055  & 0.070 \tabularnewline
\hline
\multicolumn{1}{l||}{$c=-50$} & \multicolumn{5}{c|}{$\delta=-0.95$} & \multicolumn{5}{c|}{$\delta=-0.5$} & \multicolumn{5}{c|}{$\delta=0$} & \multicolumn{5}{c|}{$\delta=0.5$} & \multicolumn{5}{c}{$\delta=0.95$}\tabularnewline
\multicolumn{1}{l||}{\textit{n}} & \multicolumn{1}{l}{T1} & \multicolumn{1}{l}{T2} & \multicolumn{1}{l}{T3} & \multicolumn{1}{l}{IVX} & \multicolumn{1}{l|}{OLS} & \multicolumn{1}{l}{T1} & \multicolumn{1}{l}{T2} & \multicolumn{1}{l}{T3} & \multicolumn{1}{l}{IVX} & \multicolumn{1}{l|}{OLS} & \multicolumn{1}{l}{T1} & \multicolumn{1}{l}{T2} & \multicolumn{1}{l}{T3} & \multicolumn{1}{l}{IVX} & \multicolumn{1}{l|}{OLS} & \multicolumn{1}{l}{T1} & \multicolumn{1}{l}{T2} & \multicolumn{1}{l}{T3} & \multicolumn{1}{l}{IVX} & \multicolumn{1}{l|}{OLS} & \multicolumn{1}{l}{T1} & \multicolumn{1}{l}{T2} & \multicolumn{1}{l}{T3} & \multicolumn{1}{l}{IVX} & \multicolumn{1}{l}{OLS}\tabularnewline
250  & 0.053  & 0.053  & 0.056  & 0.054  & 0.058  & 0.052  & 0.050  & 0.049  & 0.050  & 0.051  & 0.049  & 0.051  & 0.049  & 0.049 & 0.049  & 0.052  & 0.054  & 0.052  & 0.050  & 0.053  & 0.055  & 0.049  & 0.055  & 0.055  & 0.058 \tabularnewline
500  & 0.052  & 0.053  & 0.056  & 0.054  & 0.059  & 0.052  & 0.054  & 0.052  & 0.051  & 0.053  & 0.048  & 0.051  & 0.046  & 0.047  & 0.048  & 0.050  & 0.053  & 0.050  & 0.050  & 0.050  & 0.053  & 0.048  & 0.055  & 0.055  & 0.059 \tabularnewline
750  & 0.051  & 0.050  & 0.059  & 0.059  & 0.064  & 0.053  & 0.049  & 0.054  & 0.055  & 0.055  & 0.053  & 0.050  & 0.054  & 0.053  & 0.052  & 0.057  & 0.051  & 0.056  & 0.056  & 0.058  & 0.057  & 0.046  & 0.059  & 0.059  & 0.063 \tabularnewline
1000  & 0.054  & 0.054  & 0.058  & 0.055  & 0.061  & 0.051  & 0.053  & 0.052  & 0.053  & 0.053  & 0.050  & 0.048  & 0.048  & 0.050  & 0.050  & 0.050  & 0.047  & 0.048  & 0.049  & 0.050  & 0.051  & 0.050 & 0.054 & 0.053 & 0.058\tabularnewline
\hline
\hline
\multicolumn{1}{l}{} &  &  &  &  & \multicolumn{1}{r}{} &  &  &  &  & \multicolumn{1}{r}{} &  &  &  &  & \multicolumn{1}{r}{} &  &  &  &  & \multicolumn{1}{r}{} &  &  &  &  & \tabularnewline
\end{tabular}} \label{tab:table1}
\end{table}

\end{landscape}

\begin{figure}[H]
\caption{Empirical Power (NI regressor; 5\% Nominal Size ; $c=0$)}
\centering %
\begin{tabular}{c}
\includegraphics[width=0.9\textwidth]{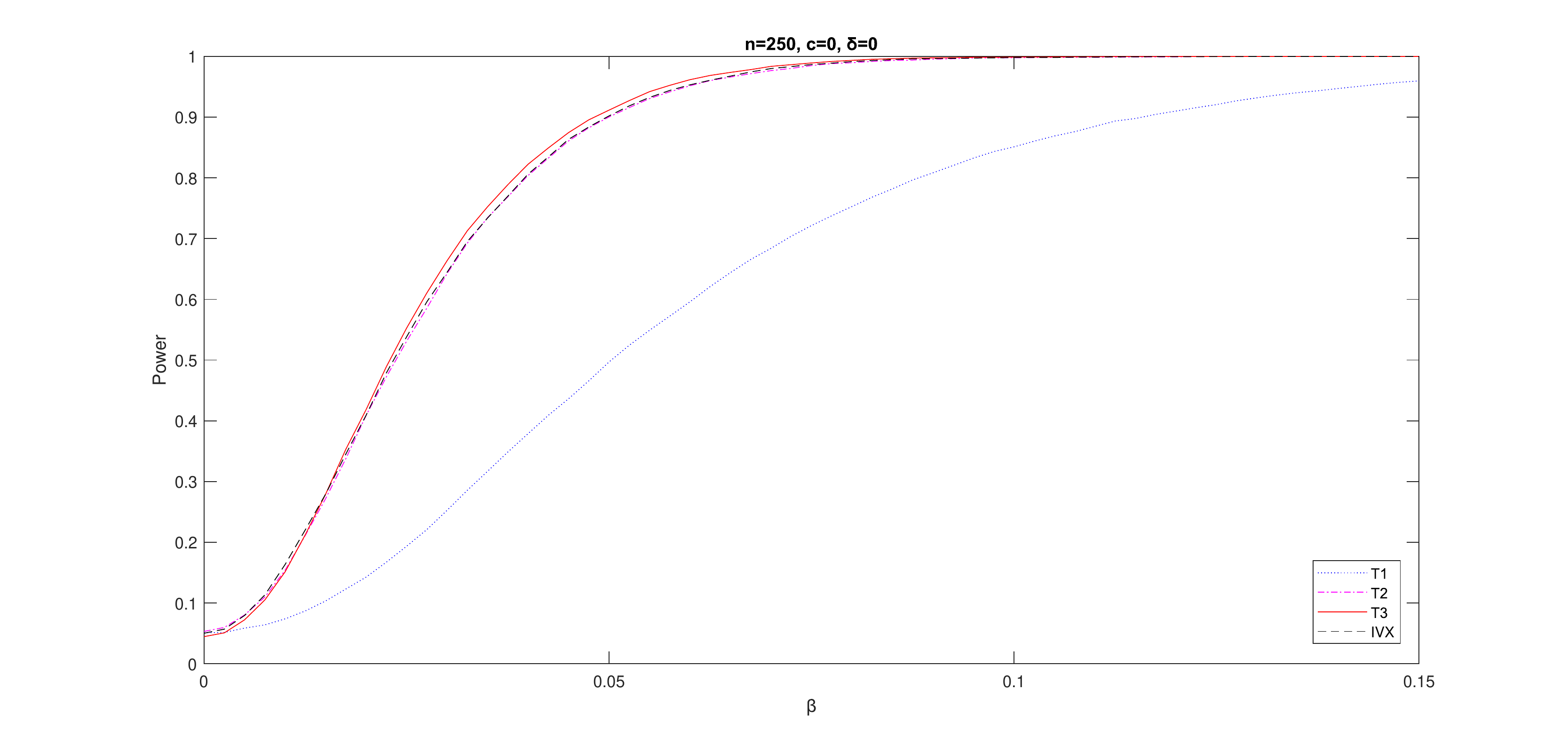} \tabularnewline
\includegraphics[width=0.9\textwidth]{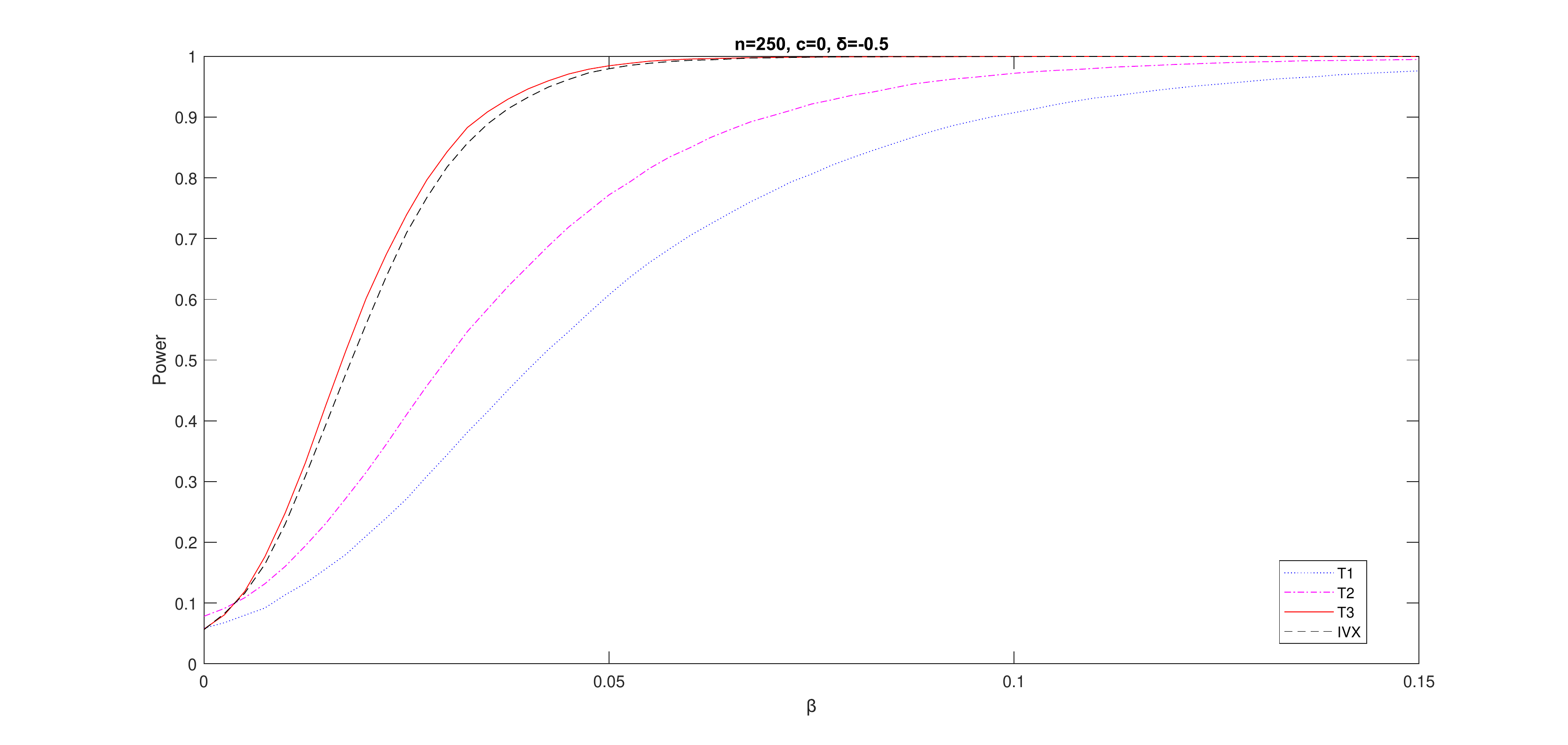} \tabularnewline
\includegraphics[width=0.9\textwidth]{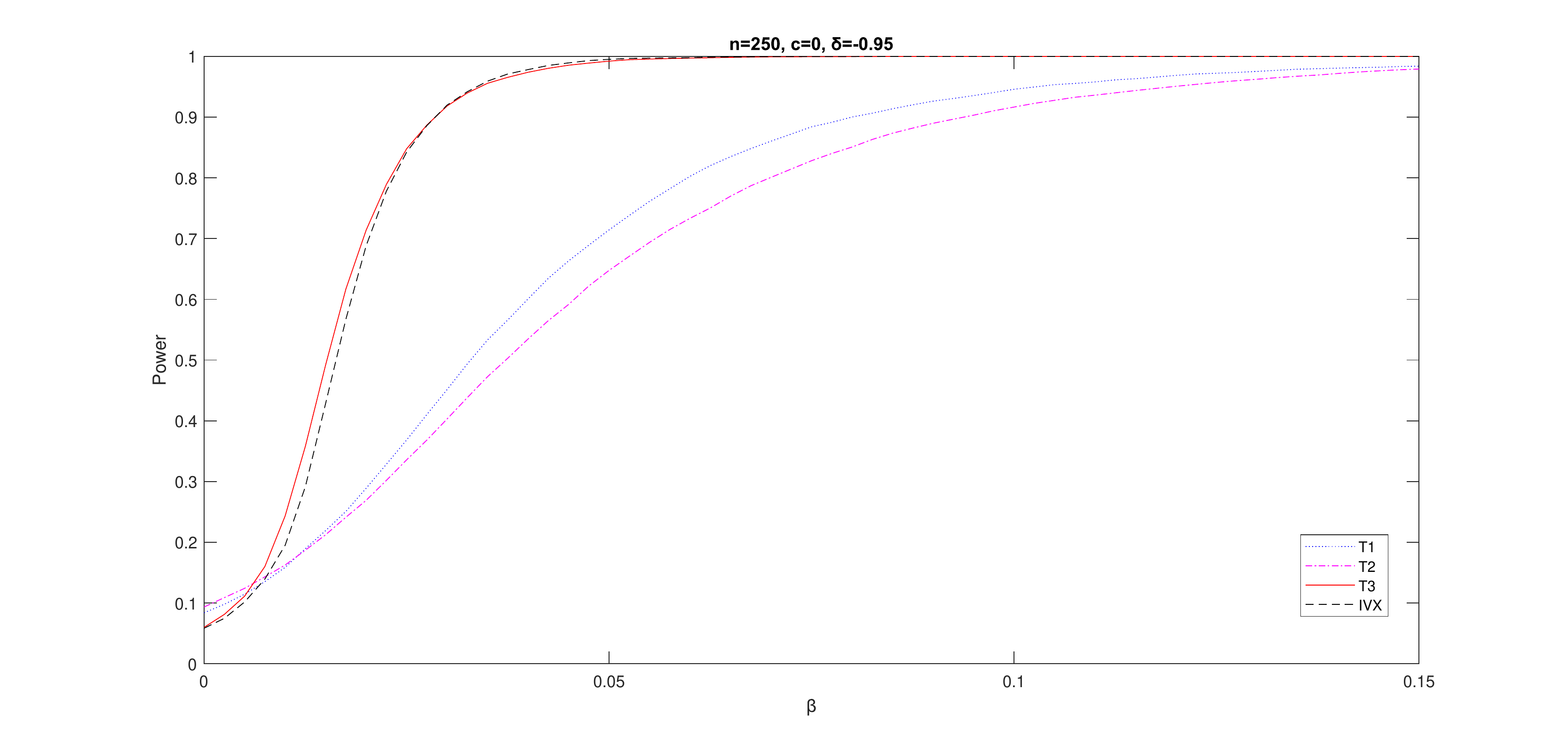}\tabularnewline
\end{tabular}\label{fig:Name}
\end{figure}

%%%%%%%%%%%%%%%%%%%%%%%%%%%%%%%%%%%%%%%%%%%%%%%%%%%%%%%%%%%%%%%%%%%%%%%%%%%%%%%%%%%%%%%%%%%%%%%%
%%%%%%%%%%%%%%%%%%%%%%%%%%%%%%%%%%%%%%%%%%%%%%%%%%%%%%%%%%%%%%%%%%%%%%%%%%%%%%%%%%%%%%%%%%%%%%%%
\begin{figure}[H]
\caption{Empirical Power (NI regressor; 5\% Nominal Size; $c=-20$)}
\centering %
\begin{tabular}{c}
\includegraphics[width=0.99\textwidth]{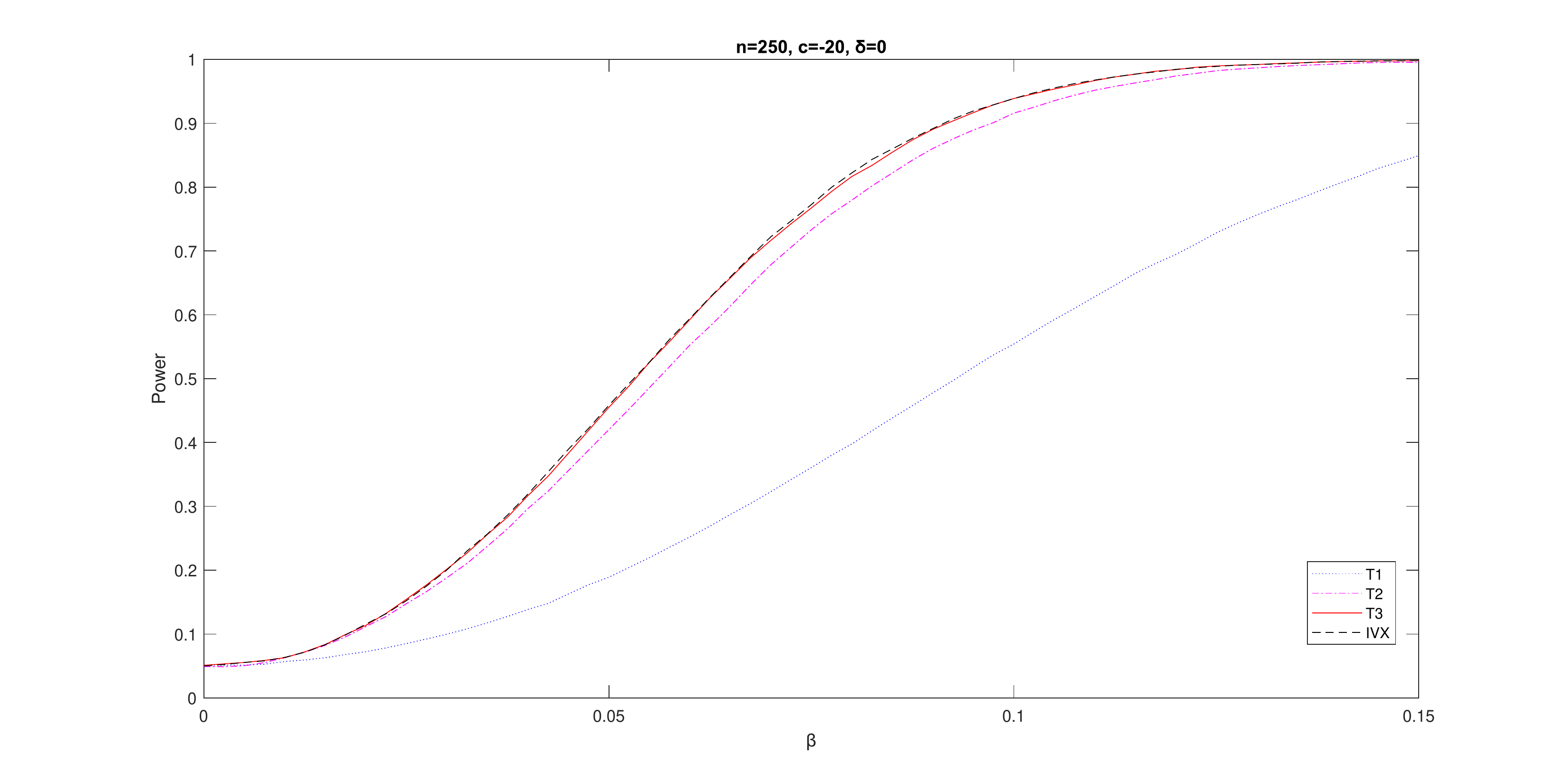}\tabularnewline
\includegraphics[width=0.99\textwidth]{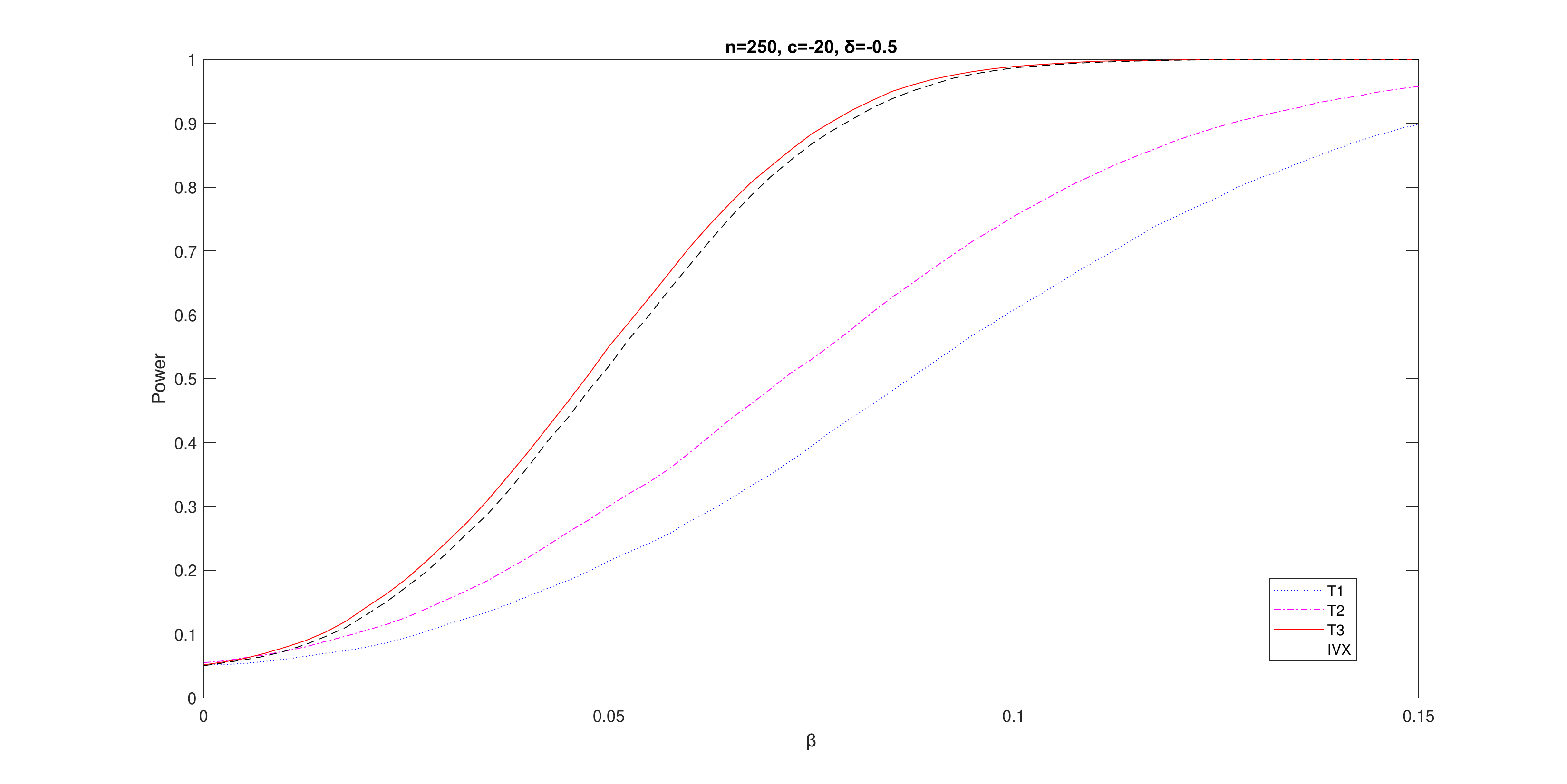}\tabularnewline
\includegraphics[width=0.99\textwidth]{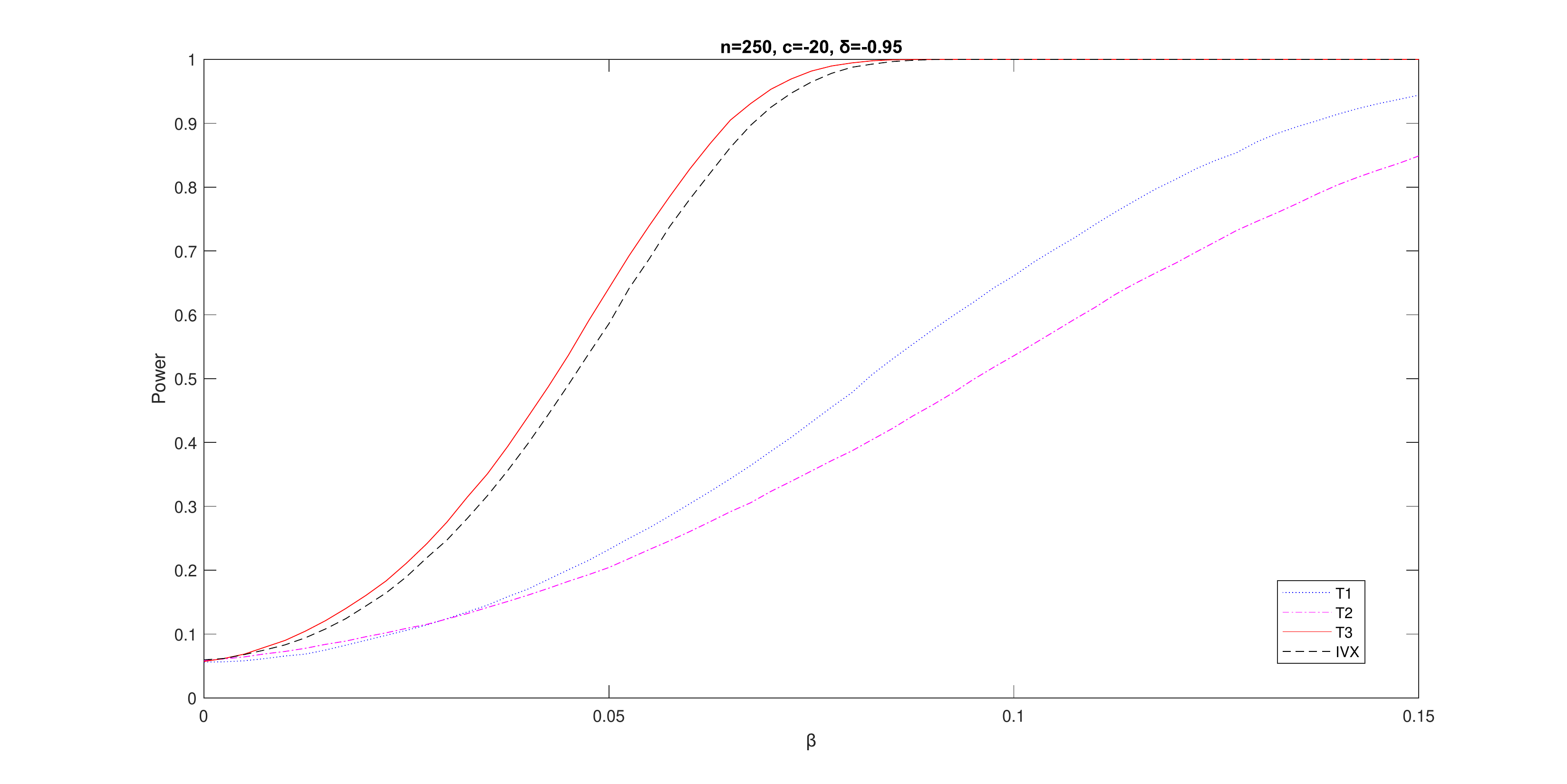}\tabularnewline
%\textbf{(d)} n=250, c=-20, $\delta$=0  & \textbf{(e)} n=250, c=-20, $\delta$=-0.5& \textbf{(f)} n=250, c=-20, $\delta$=-0.95\\[3pt]
\tabularnewline
\end{tabular}\label{fig:Name1}
\end{figure}

\begin{landscape}

\begin{table}[htbp]
\centering \caption{Empirical Size (Fractional Regressor; 5\% Nominal Size)}
\resizebox{1.05\textwidth}{!}{%
\begin{tabular}{l||rr|rr|rr|rr|rr|rr}
\multicolumn{1}{l}{} & \multicolumn{2}{c}{} & \multicolumn{2}{c}{} & \multicolumn{2}{c}{} & \multicolumn{2}{c}{} & \multicolumn{2}{c}{} & \multicolumn{2}{c}{}\tabularnewline
\hline
\hline
\multicolumn{1}{l||}{$\delta=-0.95$} & \multicolumn{2}{c|}{$d=0.75$} & \multicolumn{2}{c|}{$d=0.8$} & \multicolumn{2}{c|}{$d=0.9$} & \multicolumn{2}{c|}{$d=1$} & \multicolumn{2}{c|}{$d=1.1$} & \multicolumn{2}{c}{$d=1.2$}\tabularnewline
\multicolumn{1}{l||}{\textit{n}} & \multicolumn{1}{l}{T3} & \multicolumn{1}{l|}{LS} & \multicolumn{1}{l}{T3} & \multicolumn{1}{l|}{LS} & \multicolumn{1}{l}{T3} & \multicolumn{1}{l|}{LS} & \multicolumn{1}{l}{T3} & \multicolumn{1}{l|}{LS} & \multicolumn{1}{l}{T3} & \multicolumn{1}{l|}{LS} & \multicolumn{1}{l}{T3} & \multicolumn{1}{l}{LS}\tabularnewline
250  & 0.051  & 0.158  & 0.051  & 0.184  & 0.055  & 0.235  & 0.060  & 0.278  & 0.064  & 0.308  & 0.067  & 0.325 \tabularnewline
500  & 0.051  & 0.161  & 0.052  & 0.184  & 0.058  & 0.242 & 0.062  & 0.287  & 0.066  & 0.319  & 0.068  & 0.337 \tabularnewline
750  & 0.051  & 0.155  & 0.052  & 0.178  & 0.058  & 0.230  & 0.062  & 0.272  & 0.064  & 0.301  & 0.067  & 0.322 \tabularnewline
1000  & 0.048 & 0.155  & 0.050  & 0.183  & 0.055  & 0.229  & 0.059  & 0.278  & 0.065  & 0.310  & 0.069  & 0.327\tabularnewline
\hline
\multicolumn{1}{l||}{$\delta=-0.5$} & \multicolumn{2}{c|}{$d=0.75$} & \multicolumn{2}{c|}{$d=0.8$} & \multicolumn{2}{c|}{$d=0.9$} & \multicolumn{2}{c|}{$d=1$} & \multicolumn{2}{c|}{$d=1.1$} & \multicolumn{2}{c}{$d=1.2$}\tabularnewline
\multicolumn{1}{l||}{\textit{n}} & \multicolumn{1}{l}{T3} & \multicolumn{1}{l|}{LS} & \multicolumn{1}{l}{T3} & \multicolumn{1}{l|}{LS} & \multicolumn{1}{l}{T3} & \multicolumn{1}{l|}{LS} & \multicolumn{1}{l}{T3} & \multicolumn{1}{l|}{LS} & \multicolumn{1}{l}{T3} & \multicolumn{1}{l|}{LS} & \multicolumn{1}{l}{T3} & \multicolumn{1}{l}{LS}\tabularnewline
250  & 0.051  & 0.085  & 0.052  & 0.093  & 0.055  & 0.107  & 0.057  & 0.117  & 0.058  & 0.121  & 0.057  & 0.126 \tabularnewline
500  & 0.051  & 0.085  & 0.050  & 0.091  & 0.051  & 0.102  & 0.051  & 0.114  & 0.052  & 0.120  & 0.052  & 0.123 \tabularnewline
750  & 0.048  & 0.081  & 0.046  & 0.086  & 0.048  & 0.098  & 0.051  & 0.109  & 0.052  & 0.117  & 0.052  & 0.119 \tabularnewline
1000  & 0.047  & 0.077  & 0.047  & 0.086  & 0.048  & 0.102  & 0.051  & 0.111  & 0.056  & 0.118  & 0.053  & 0.120 \tabularnewline
\hline
\multicolumn{1}{l||}{$\delta=0.0$} & \multicolumn{2}{c|}{$d=0.75$} & \multicolumn{2}{c|}{$d=0.8$} & \multicolumn{2}{c|}{$d=0.9$} & \multicolumn{2}{c|}{$d=1$} & \multicolumn{2}{c|}{$d=1.1$} & \multicolumn{2}{c}{$d=1.2$}\tabularnewline
\multicolumn{1}{l||}{\textit{n}} & \multicolumn{1}{l}{T3} & \multicolumn{1}{l|}{LS} & \multicolumn{1}{l}{T3} & \multicolumn{1}{l|}{LS} & \multicolumn{1}{l}{T3} & \multicolumn{1}{l|}{LS} & \multicolumn{1}{l}{T3} & \multicolumn{1}{l|}{LS} & \multicolumn{1}{l}{T3} & \multicolumn{1}{l|}{LS} & \multicolumn{1}{l}{T3} & \multicolumn{1}{l}{LS}\tabularnewline
250  & 0.043  & 0.051  & 0.044  & 0.052  & 0.043  & 0.053  & 0.045  & 0.053  & 0.044  & 0.052  & 0.043  & 0.052 \tabularnewline
500 & 0.048  & 0.054  & 0.048  & 0.055  & 0.047  & 0.054  & 0.046  & 0.054  & 0.045  & 0.055  & 0.045  & 0.056 \tabularnewline
750 & 0.048  & 0.053  & 0.046  & 0.051  & 0.043  & 0.052  & 0.042  & 0.051  & 0.042  & 0.051  & 0.043  & 0.053 \tabularnewline
1000  & 0.045  & 0.049  & 0.044  & 0.049  & 0.044  & 0.053  & 0.046  & 0.053  & 0.044  & 0.054  & 0.044  & 0.055 \tabularnewline
\hline
\hline
\multicolumn{1}{l}{} &  & \multicolumn{1}{r}{} &  & \multicolumn{1}{r}{} &  & \multicolumn{1}{r}{} &  & \multicolumn{1}{r}{} &  & \multicolumn{1}{r}{} &  & \tabularnewline
\end{tabular}} \label{tab:table2}
\end{table}

\end{landscape}
\begin{figure}[H]
\caption{Empirical Power (Fractional Regressor; 5\% Nominal Size)}
\centering %
\begin{tabular}{c}
\includegraphics[width=0.9\textwidth]{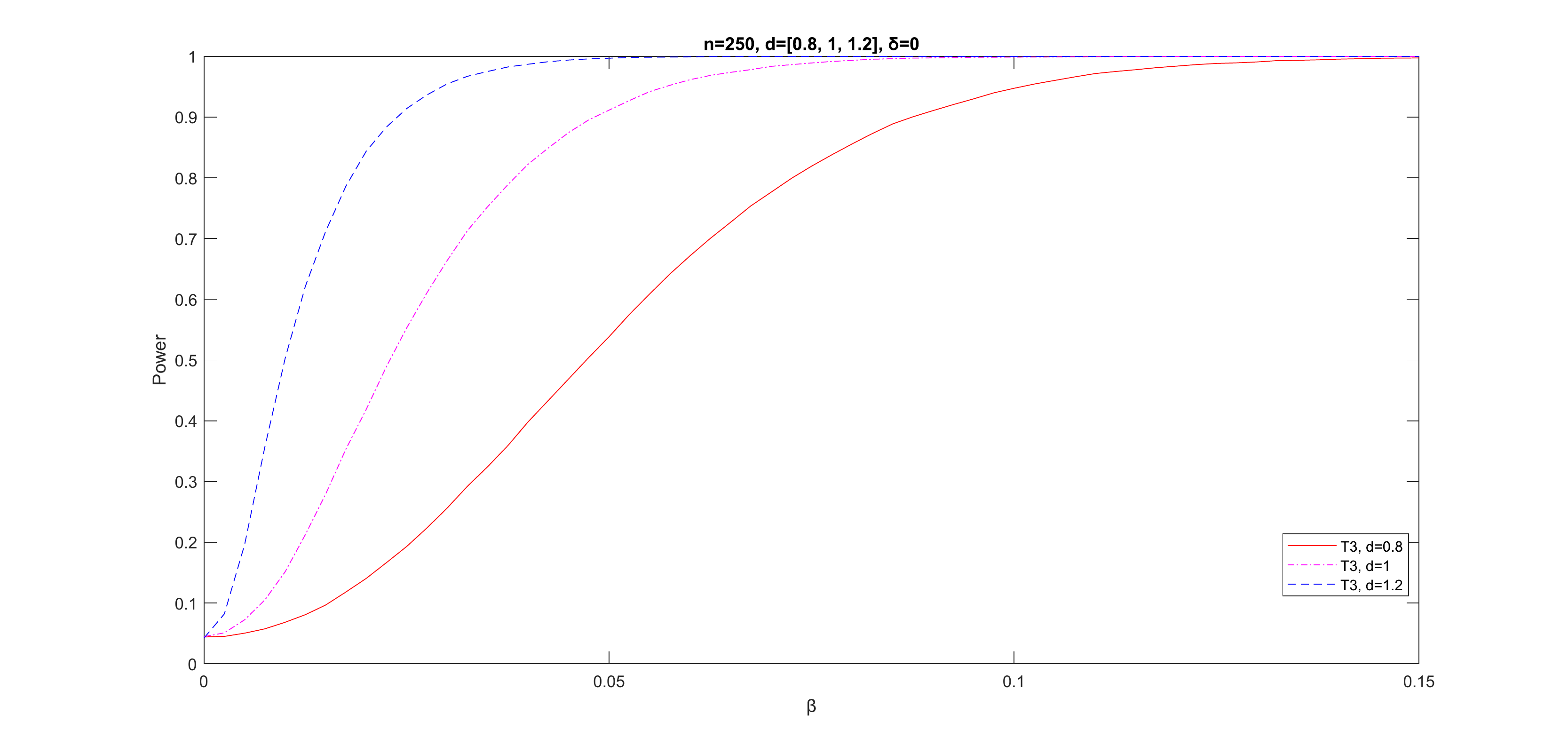} \tabularnewline
\includegraphics[width=0.9\textwidth]{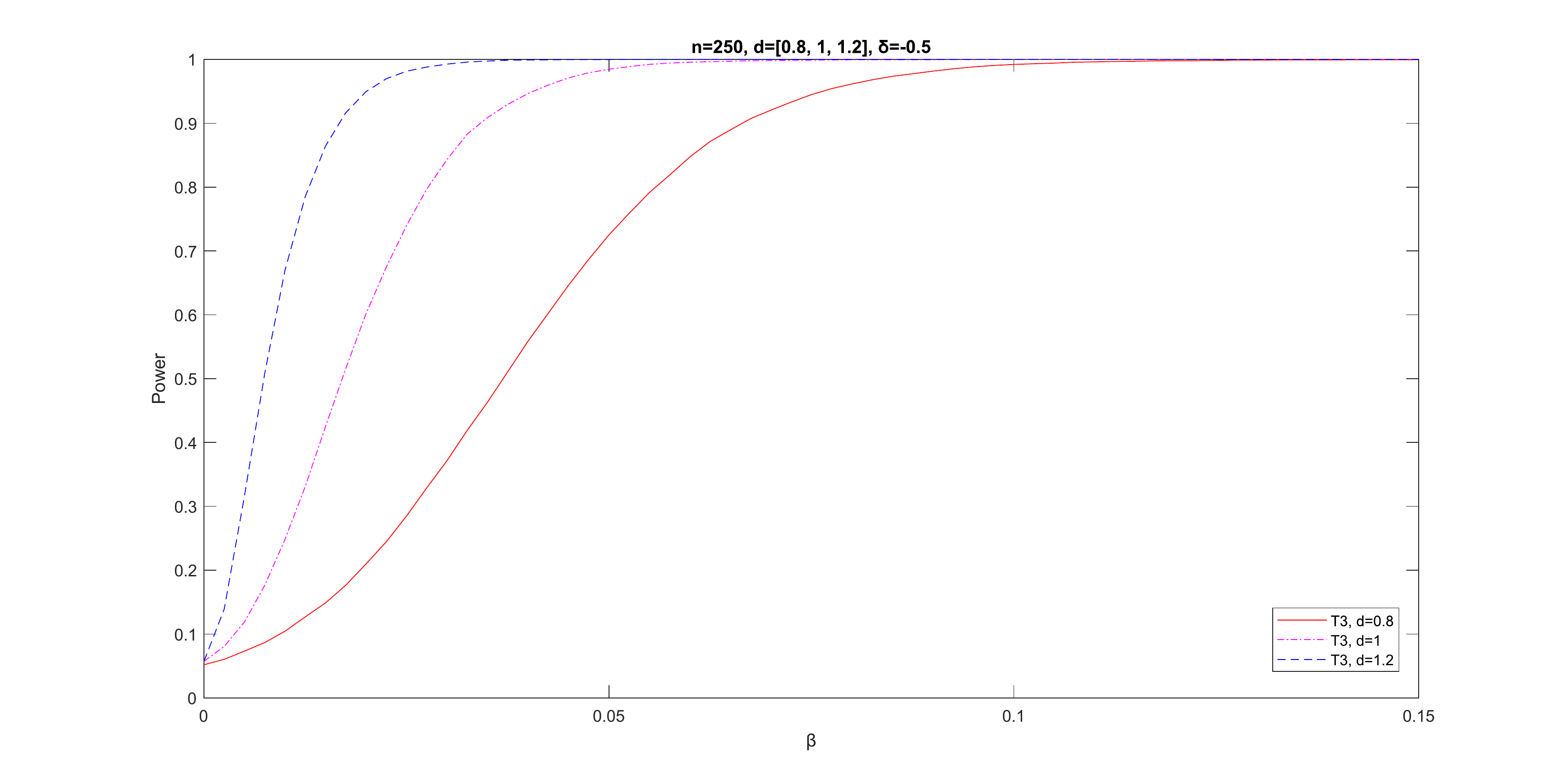} \tabularnewline
\includegraphics[width=0.9\textwidth]{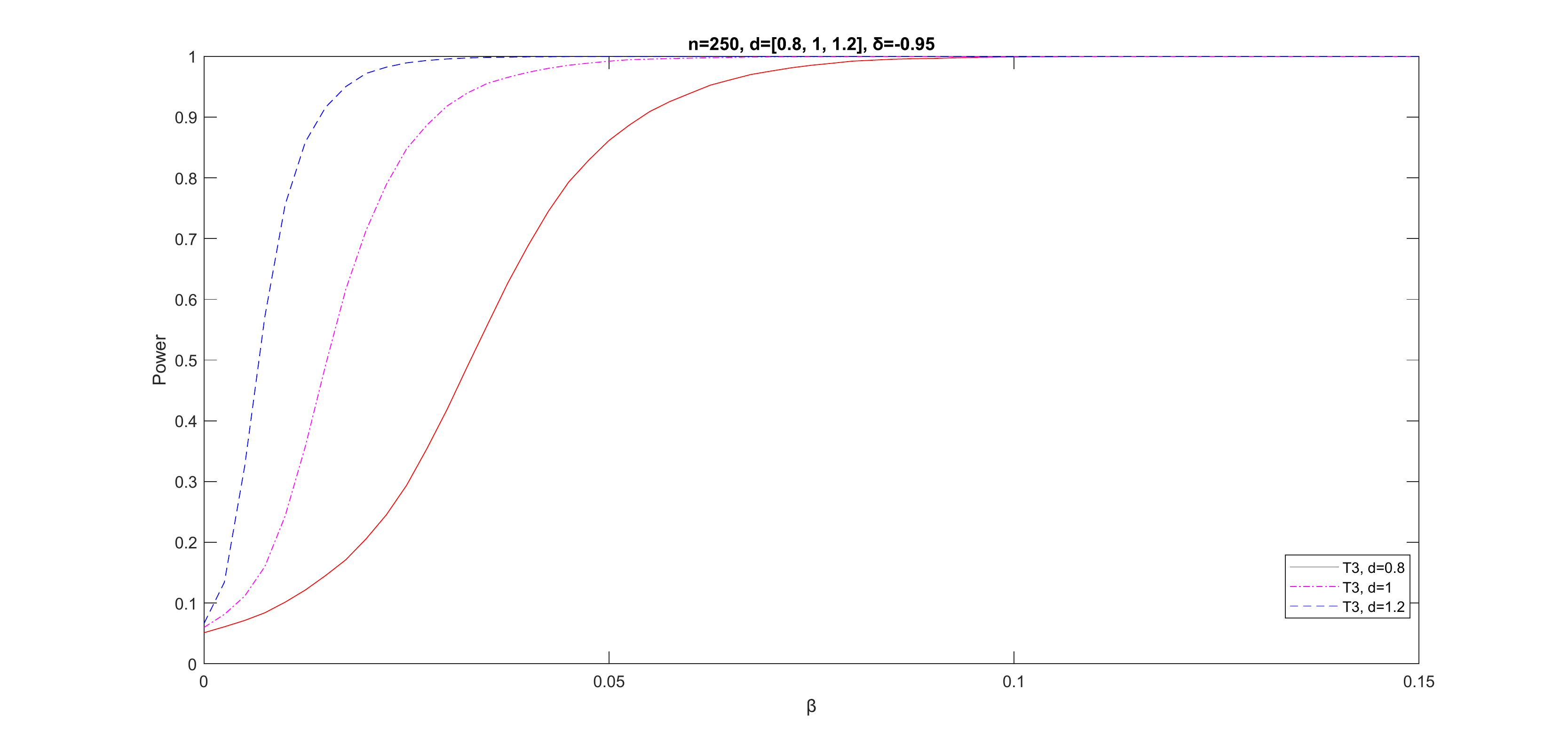}\tabularnewline
%\textbf{(a)} n=250,  $\delta$=0  & \textbf{(b)} n=250, $\delta$=-0.5 & \textbf{(c)}  n=250,  $\delta$=-0.95  \\[3pt]
\tabularnewline
\end{tabular}\label{fig:Name3}
\end{figure}

%%%%%%%%%%%%%%%%%%%%%%%%%%%%%%%%%%%%%%%%%%%%%%%%%%%%%%%%%%%%%%%%%%%%%%%%%%%%%%%%%%%%%%%%%%%%%%%%

\section{Application to the predictability of stock returns}

A large literature in empirical finance is devoted to the investigation
of the hypothesis that stock returns can be predicted with publicly
available information. For a review of existing work see for example
Welch and Goyal (2008) and for more recent developments Kostakis,
Magdalinos and Stamatogiannis (2015). Typically empirical work in
this area involves inferential procedures, for the hypothesis $H_{0}:\beta=0$,
in the context of \textit{predictive} regressions of the form
\begin{equation}
r_{k+1}=\mu+\beta x_{k}+u_{k+1},\label{App1}
\end{equation}
where $r_{k}$ are stock returns relating to some stock index,$\ x_{k}$
some predictive variable and $u_{t}$ a martingale difference regression
error. Usually some financial ratio (e.g. dividend yield, earnings
to price ratio, book to market ratio) or some macroeconomic variable
(e.g. inflation) is considered as a possible predictor for future
returns. Phillips (2015) provides a review for the econometric methodology
employed in the predictive regressions literature. Most studies (e.g.
Welch and Goyal, 2008) are utilising methods that are only valid for
stationary $x_{k}$ despite the fact that there is strong evidence
that that in certain datasets various financial and macroeconomic
variables are consistent with nonstationary processes (e.g. see Kostakis
et al, 2015; Table 4). To the best of our knowledge, Campbell and
Yogo (2006)\ is the first work that explicitly provides an attempt
to address the possibility that the regressor is nonstatationary.
In particular, Campbell and Yogo (2006) develop a testing procedure
for the case the predictor is a NI array based on conservative confidence
intervals. Kostakis et al (2015) consider a modified version of the
Magdalinos and Phillips (2009) IVX, that involves a finite sample
correction relating to intercept estimation, to examine the return
predictability hypothesis. The IVX estimator yields conventional inference
for the case where $x_{k}$ is a NI or mildly integrated array (e.g.
Phillips and Magdalinos, 2007) or a stationary linear process. IVX
instruments are also employed in the recent work of Demetrescu et
al (2020) who propose inferential procedures for detecting episodic
predictability in stock returns. The IVX method has been also employed
in the recent work of Yang, Long, Peng and Cai (2020) who investigate
predictability in the U.S. housing index return.

An important issue that has been largely overlooked in most studies
in this area, is that stock returns series typically exhibit very
weak persistence relative to most popular predictors. In particular,
in many datasets short-term returns appear to be close to $I(d)$
processes with $d\approx0$, whilst several predictors appear to be
$I(d)$ with $d>1/2$ i.e. nonstationary processes. Regressing a stationary
processes on a possibly nonstationary leads to \textit{misbalancing}.
As emphasised by Phillips (2015), misbalancing may result to asymptotically vanishing estimators. For instance if $r_{k}\sim I(d)$ with
$d<1/2$ (stationary long memory) and $x_{k}\sim I(d)$ with $d>1/2$,
then then OLS estimator for $\beta$ in (\ref{App1}) is $\tilde{\beta}\rightarrow_{P}0$.

Only a few studies in this area attempt to address the issue of misbalancing.\ Marmer
(2007) points out that a nonlinear relationship between returns and
predictive variables is a plausible justification for this discrepancy
in persistence. It is known for instance that integrable and bounded
transformations of persistent processes may exhibit very weak signal
(e.g. Park and Phillips, 1999, 2001; Park, 2003). Therefore, suppose
that $r_{k+1}=f\left(x_{k}\right)+u_{k+1}$ where $f$ is either integrable
and compactly supported or the indicator function $1\left\{ .<0\right\} $.
The predictor $x_{k}$ in this case has only ``spatial episodic''\ impact
on returns when the predictive variable visits the support of $f$
(integrable case) or when it assumes negative values (indicator case).\ For
DGPs of this kind it is difficult distinguishing $r_{k}$ from the
martingale difference error $u_{k}$, despite the fact $r_{k}$ is
a function of a persistent process\ (see for example Kasparis, Andreou
and Phillips (2015), Figure 6; or Phillips (2015), Figure 2). Marmer
(2007) develops a RESET type of functional form test for detecting
possibly nonlinear components (e.g. integrable) of some predictor
in the stock return series. A similar approach is also followed by
Kasparis (2010) and Kasparis et al (2015), who utilise test statistics
that involve integrable transformations of the predictor. The presence
of integrable transformations in the test statistics results in conventional
inference but can also detect weak signal nonlinear components affecting
the returns series (for more details see p. 473-474 in Kasparis et
al, 2015). Bollerslev, Osterrieder, Sizova and\ Tauchen (2013) follow
a different approach for addressing the issue of misbalancing. These
authors consider \textit{vix} and \textit{realised volatility} as
possible predictors of stock returns. Using preliminary estimations
they find that the aforementioned predictors exhibit long memory with
memory parameters $d\approx0.4$, whilst stock returns appear to have
a memory parameter $d\approx0$. In view of this, Bollerslev et al
(2013) consider prefiltered predictors of the form $\left(I-L\right)^{\hat{d}}x_{k}$
where $x_{k}$\ is some volatility variable. Notice that the fractionally
differenced process is approximately $d\approx0$. Finally, Demetrescu
et al (2020) develop inferential procedures capable of detecting episodic
predictability is stock returns for the case where the predictors
that are either $I(0)$ or NI. In particular, they consider a potentially
nonlinear relationship between returns and the predictive variables
of the form $r_{k+1}=f_{n}\left(x_{k},k/n\right)+u_{k+1}$, where
$f_{n}\left(x_{k},k/n\right)=\mu+k_{n}\beta\left(k/n\right)x_{k}$,
$\beta\left(.\right)$ is a TVP depending on the rescaled time trend
$k/n$, and $k_{n}$ an appropriate sequence. This formulations allows for ``time episodic''\ impact of the predictor to the returns variable. Demetrescu et al (2020)
achieve conventional inference by either utilising IVX instruments
or the so called type II instruments of Breitung and Demetrescu (2015).\footnote{The method of Demetrescu et al (2020) can be used in conjunction with
various instruments including LTLS. Such a development would require
additional theoretical work and is left for future research.}

In this work we address the issue of misbalancing by consider predictability
over longer horizons. In particular, we employ LTLS based inference
in predictive regressions of the form
\begin{equation}
r_{k+m}=\mu+\beta x_{k}+u_{k+m},\label{App2}
\end{equation}
where $m\geq1$. The specification of (\ref{App2}) has been considered
by other studies that investigate return predictability over long
horizons (see for example Bandi and Perron, 2008; Hjalmarsson, 2011).
The data are taken from the updated 2018 Welch and Goyal dataset\footnote{The data are download from Amit Goyal's webpage: http://www.hec.unil.ch/agoyal/

{}}. The returns variable is constructed from the SP500 index ($I_{k}$)
as follows $r_{k+m}=\ln(I_{k+m})-\ln(I_{k})$. We are using monthly
and quarterly observations. Therefore, for monthly data, $r_{k+m}$
should be understood as $m$ months ahead returns, and for quarterly
observations as $m$ quarters ahead. By construction returns are log-price
differences. Therefore, the persistence of the returns series tends
to increase as the horizon increases. Table 3 provides memory estimates
for the return series over different horizons and frequencies. In
particular, we use the local Whittle estimator (LW; e.g. see Robinson,
1995) and the exact local Whittle (ELW) of Shimotsu and Phillips (2005).
The bandwidth employed is of the form $n^{b}$. Shimotsu and Phillips
(2005) consider $b=0.65$ for the bandwidth exponent. Here we also
consider $b=0.55$ and $b=0.75$. Moreover, we report memory estimates
for the earnings to price ratio (EP). The particular series appears
to be less persistent than dividend yield and book to marker ratio
that are commonly used in empirical work. For this reason we will
concentrate on EP whose memory characteristics are closer to those
of the returns series. It can be seen from Table 3 that the EP appears
to be nonstationary at both frequencies and for all bandwidth choices
with minimal memory estimate $0.76$. Further, the memory characteristics of the returns series appear to resemble those of the EP variable over longer horizons i.e. $m=24$ for monthly data and $m=12$ for quarterly, when $b=0.65, 0.75$. 

\begin{table}
\caption{Memory Estimates}
\centering %
\begin{tabular}{l||ll|ll|ll}
\multicolumn{7}{c}{}\tabularnewline
\hline
\hline
\multicolumn{7}{c}{Monthly Data}\tabularnewline
\hline
Bandwidth $n^{b}$ & \multicolumn{2}{c|}{$b=0.55$} & \multicolumn{2}{c|}{$b=0.65$} & \multicolumn{2}{c}{$b=0.75$}\tabularnewline
 & LW & ELW & LW & ELW & LW & ELW\tabularnewline
Returns ($m=1$) & -0.09 & -0.08 & 0.07 & 0.06 & 0.03 & 0.04\tabularnewline
Returns ($m=12$) & -0.036 & -0.02 & 0.45 & 0.45 & 0.84 & 0.86\tabularnewline
Returns ($m=24$) & 0.21 & 0.22 & 0.93 & 0.93 & 1.04 & 1.06\tabularnewline
EP & 0.77 & 0.85 & 0.92 & 1.22 & 1.02 & 1.51\tabularnewline
\hline
\multicolumn{7}{c}{Quarterly Data}\tabularnewline
\hline
Bandwidth $n^{b}$ & \multicolumn{2}{c|}{$b=0.55$} & \multicolumn{2}{c|}{$b=0.65$} & \multicolumn{2}{c}{$b=0.75$}\tabularnewline
 & LW & ELW & LW & ELW & LW & ELW\tabularnewline
Returns ($m=1$) & -0.09 & -0.07 & -0.09 & -0.08 & 0.03 & 0.04\tabularnewline
Returns ($m=8$) & -0.03 & -0.01 & 0.16 & 0.17 & 0.89 & 0.93\tabularnewline
Returns ($m=12$) & 0.06 & 0.08 & 0.82 & 0.83 & 1.19 & 1.14\tabularnewline
EP & 0.76 & 0.81 & 0.79 & 0.85 & 0.88 & 1.17\tabularnewline
\hline
\hline
\multicolumn{1}{l}{} &  & \multicolumn{1}{l}{} &  & \multicolumn{1}{l}{} &  & \tabularnewline
\end{tabular}
\end{table}

%%%%%%%%%%%%%%%%%%%%%%%%%%%%%%%%%%%%%%%%%%%%%%%%%

Figure 3 reports values for the LTLS $\hat{T}$-statistics for the
hypothesis $H_{0}:\beta=0$ vs $H_{1}:\beta\neq0$ (c.f. equation
(\ref{App2})). These values are plotted against the predictability
horizon parameter $m$. We consider three configurations for kernels,
\textit{cps} and bandwidth sequences consistent with the set-ups S1,
S2 and S3 given in the previous section. In particular, for S1 and
S2 we choose $K(x)=\varphi_{0.1\tilde{\sigma}_{u}^{2}}(x)^{1/2}$,
$K(x)^{\ast}=\varphi_{\tilde{\sigma}_{u}^{2}}(x)^{1/2}$. It can be
seen from Figure 3 that there is evidence for predictability only
for longer horizons under S1 and S3. For monthly data,
the null hypothesis is rejected at a 5\%\ level under S1 and S3 for
for $m$ greater than 6 and 5 respectively. For quarterly data the
null is rejected under S1 and S3 for   $m$ greater than 12 and
10 respectively. These findings are consistent with those of Bandi
and Perron (2008) how find strong predictability (by volatility predictors)
over longer horizons.

\begin{figure}[H]
\caption{Predictability Tests}
\centering \hspace*{-2.5cm} %
\begin{tabular}{c}
\includegraphics[width=1.3\textwidth]{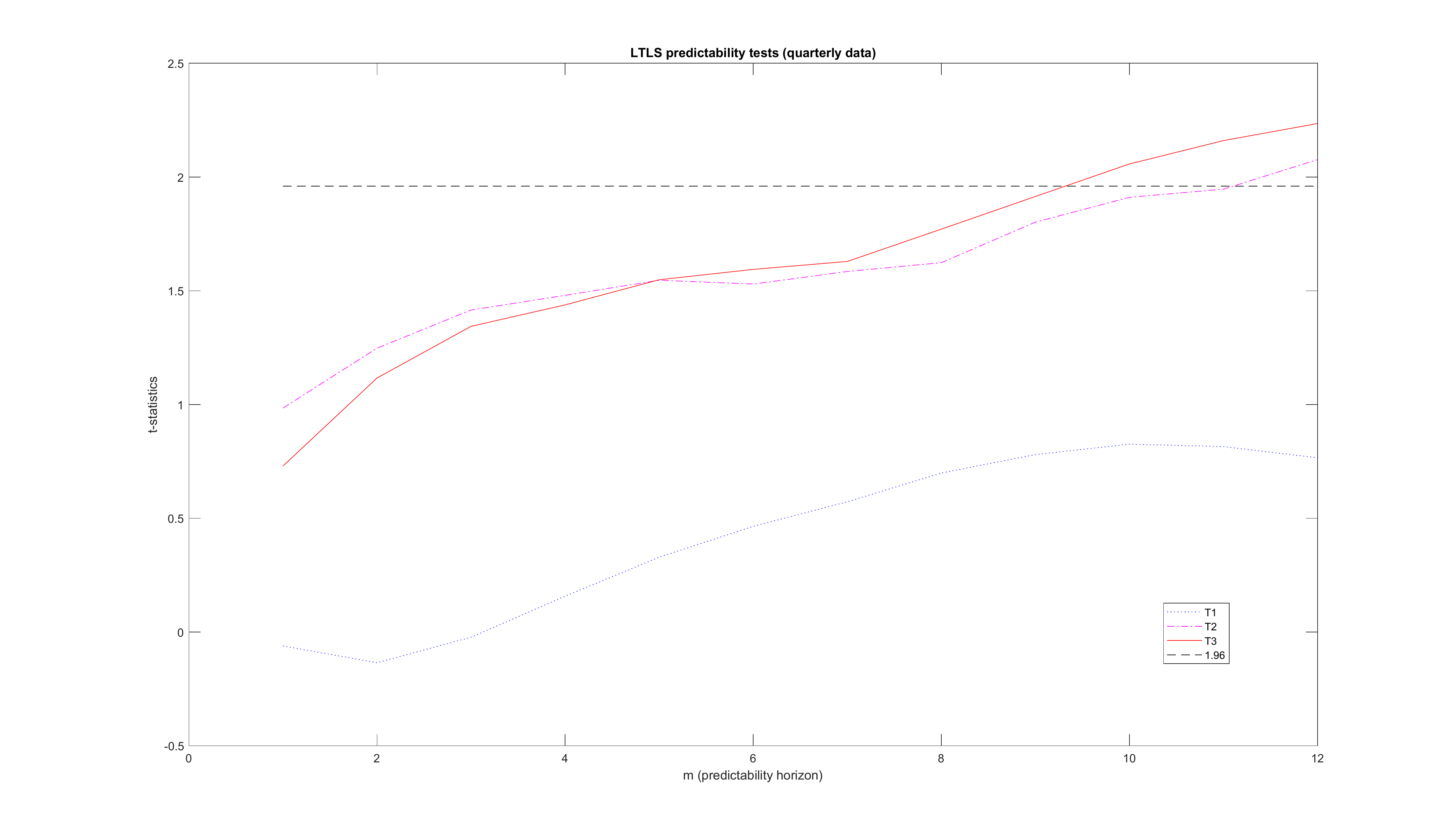}\tabularnewline
\includegraphics[width=1.3\textwidth]{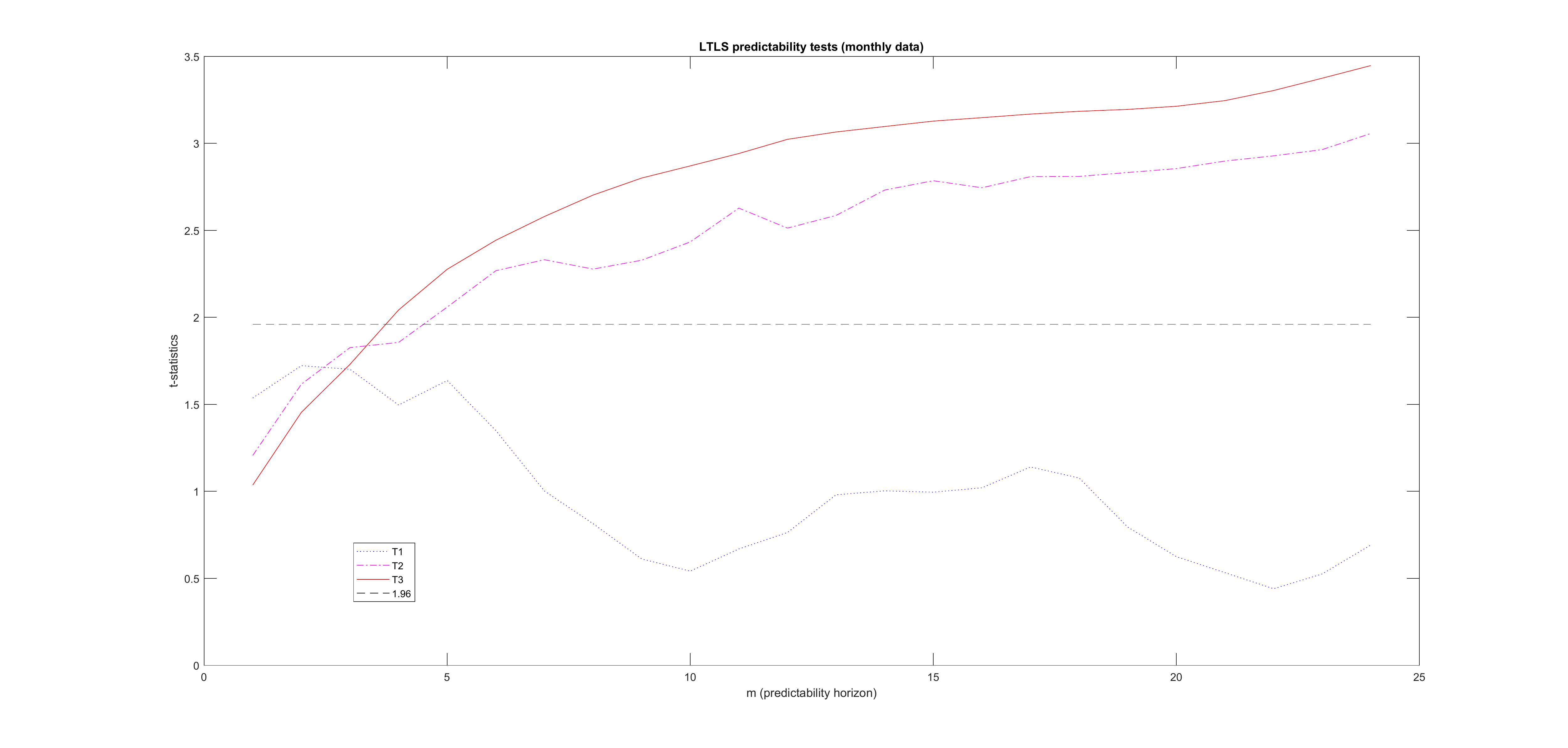} \tabularnewline
%\textbf{(a)} LTLS predictability tests (quarterly data) & \textbf{(b)} LTLS predictability tests (monthly data) \\[3pt]
\tabularnewline
\end{tabular}\label{fig:Name4}
\end{figure}

%%%%%%%%%%%%%%%%%%%%%%%%%%%%%%%%%%%

\section{Proofs of main results}

Throughout the section, we assume that  $C, C_0, C_1, C_2,...$ are positive  constants
that may take a different value in each appearance and let $K_{kn}:=\sum_{j=1}^{l_{n}}K\left[c_{n}(k/n-\tau_{j})\right]$ as in (\ref {kernel1}).

\subsection{Preliminaries}

We start with two preliminary lemmas, which provide  significant extension to Lemma
4.1 of  Hu, Phillips and Wang (2019) and include (\ref {Lemma1a}) and (\ref {Lemma2a}) as a corollary. The proofs of these two lemmas  will be given in Sections 6.7 and 6.8, respectively.

Let $\{X_{n, k}\}_{k\geq 1,n\geq 1}$, where $X_{n, k}=(X_{nk,1},...,X_{nk,p})$,
be a vector random array. When there is no confusion, we also use the notation $X_{nk}=X_{n,k}.$
 Let $\{v_{k}\}_{k\ge 1}$ be a sequence of random variables, and $G(q)=G(q_{1},...,q_{p})$ and $K(x)$ be Borel functions on $%
%TCIMACRO{\U{211d} }%
%BeginExpansion
\mathbb{R}
%EndExpansion
^{p}$ and $%
%TCIMACRO{\U{211d} }%
%BeginExpansion
\mathbb{R}
%EndExpansion
$, respectively. For $0<\tau _{1}<\tau _{2}<...<\tau _{l}<1$, set
\begin{equation*}
S_{n,l}=\frac{c_{n}}{n}\sum_{k=1}^{n}G(X_{nk})\,v_{k}\,\frac{1}{l}%
\sum_{j=1}^{l}K\big[c_{n}(k/n-\tau _{j})\big],
\end{equation*}%
where $\{c_{n}\}_{n\ge 1}$ is a sequence of positive constants. Our first result
investigates the asymptotics of $S_{n,l}$.

\begin{lem}
\label{thm27} Suppose that

\begin{itemize}
\item[(a)] there is a continuous limiting process $%
X_{t}=(X_{1}(t),...,X_{p}(t))$ such that $X_{n,[nt]}\Rightarrow X_{t}$ on $%
D_{%
%TCIMACRO{\U{211d} }%
%BeginExpansion
\mathbb{R}
%EndExpansion
^{p}}[0,1]$;

\item[(b)] $\sup_{k\geq 1}E|v_{k}|<\infty $ and there exist $A_{0}\in
\mathbb{R}$ and $0<m:=m_{n}\rightarrow \infty $ satisfying $n/m\rightarrow
\infty $ so that $\max_{m\leq j\leq n-m}E\,\big|\frac{1}{m}%
\sum_{k=j+1}^{j+m}v_{k}-A_{0}\big|=o(1);$

\item[(c)] $G(q)$ is continuous, $K(x)$ has compact support or  $K(x)$ is eventually monotonic with
$K(x)\le C/(1+|x|)$,  and $K(x)\geq 0$ satisfying $%
\int K<\infty $.
\end{itemize}
Then, for any fixed $l\geq 1$, $c_{n}\rightarrow \infty $ and $%
c_{n}/n\rightarrow 0$, we have
\be
S_{n,l} &=&\frac{1}{l}\sum_{j=1}^{l}G(X_{n,[n\tau _{j}]})\,A_{0}\,\int
K+o_{P}(1)  \notag \\
&\rightarrow _{p} &\frac{1}{l}\sum_{j=1}^{l}G(X_{\tau _{j}})\,A_{0}\,\int K.
\label{mai}
\ee
If in addition $\tau _{j}=j/(l_{n}+1), j=1,2,...,l_{n}$, where $l_n^{-1}+l_n/c_n \to 0$, then
\be
S_{n,l_{n}}=\int_{0}^{1}G(X_{n,[nt]})dt\,A_{0}\,\int K+o_{P}(1)\rightarrow
_{p}\int_{0}^{1}G(X_{t})dt\,A_{0}\,\int K. \label{mai1}
\ee
\end{lem}

\begin{rem}
Weak convergence in (a) and  continuity of $%
G(q)$ are essentially necessary for this kind of result. The result can be
extended to the case that $G(q)$ is locally Lebesgue integrable if we impose
more smooth conditions on $X_{nk}$, but it involves more complicated
calculation. We do not pursue the extension to keep this paper under
reasonable length. It is worth to mention that no relationship is imposed
between $v_{k}$ and $X_{nk}$ and condition (b) is satisfied with $%
A_{0}=Ev_{1}$ whenever $v_{t}$ is ergodic (strictly) stationary satisfying $%
E|v_{1}|<\infty $ and $\frac{1}{n}\sum_{k=1}^{n}v_{k}\rightarrow
_{L_{1}}Ev_{1}$.
\end{rem}

If we are only interested in the boundedness of $S_{n, l}$, condition (b)
can be reduced as seen in the following result.

\begin{lem}\la {thm27a} Suppose that  conditions (a) and (c) of Lemma \ref {thm27} hold  and $\{v_k\}_{k\ge 1}$ is an arbitrary  random sequence satisfying $\sup_{k\ge 1}E|v_k|<\infty$.  Then, for any  $l\ge 1$ (allowing for $l=l_n\to \infty$), $c_n\to\infty$ and $c_n/n\to 0$, we have
\be
 \frac {c_n}n  \sum_{k=1}^n ||G( X_{nk})||\, |v_k|\, \frac{1}{l}\sum_{j=1}^{l}K\big[c_n (k/n-\tau_j)\big] = O_P(1). \la {mai9}
\ee
If  in addition $K(x)\le C_0/(1+|x|)$,  $\tau_j=j/(l_n+1), j=1,2,..., l_n$,\, $l_n\log l_n/ c_n\to 0$ and $l_n\to\infty$,  then
\be
 \frac {c_n}n  \sum_{k=1}^n ||G( X_{nk})||\, |v_k|\, \frac{1}{l_n}\sum_{1\le i<j\le l_n}K\big[c_n (k/n-\tau_i)\big]K\big[c_n (k/n-\tau_j)\big] &= & o_P(1), \la {mai9a}\\
 \frac {c_n}n  \sum_{k=1}^n ||G( X_{nk})||\, |v_k|\,\Big( \frac{1}{\sqrt {l_n}}\,\sum_{j=1}^{l_n}K\big[c_n (k/n-\tau_j)\big]\Big)^2&= & O_P(1), \la {mai9a1}\\
 \big( \frac {c_n}n\big)^2  \sum_{k=1}^n ||G( X_{nk})||\, |v_k|\, \Big(\frac{1}{ \sqrt {l_n}}\,\sum_{j=1}^{l_n}K\big[c_n (k/n-\tau_j)\big]\Big)^4&= & o_P(1). \la {mai9b}
\ee
\end{lem}

\begin{rem} Let $K^{\ast}(x)$ be  an another  positive function satisfying the same condition as that of $K(x)$. The same argument as in the proof of (\ref {mai9a}) yields
\be
 \frac {c_n}n  \sum_{k=1}^n ||G( X_{nk})||\, |v_k|\, \frac{1}{l_n}\sum_{1\le i<j\le l_n}K\big[c_n (k/n-\tau_i)\big]K^{\ast}\big[c_n (k/n-\tau_j)\big] &= & o_P(1). \la {adt1}
\ee
This, together with Lemma  \ref {thm27}, implies that
 \be
&& \frac{c_{n}}{n}\sum_{k=1}^{n}G(X_{nk})\, v_k\,\frac{1}{l_{n}}\sum_{j=1}^{l_{n}}K\left[c_{n}(k/n-\tau_{j})\right]
\sum_{j=1}^{l_{n}}K^{\ast}\left[c_{n}(k/n-\tau_{j})\right] \no\\
&=& \int_0^1 G(X_{n, [nt]}) dt\, A_0\, \int KK^{\ast}+o_P(1) \rightarrow_{d} \int_0^1G(X_t)dt \, A_0\,\int KK^{\ast}.\label{double2}
\ee
Let $0<\tau^{\ast}<1$, and $\tau^*\not \in \{\tau_j, j=1,\dots, l\}$ if $l_n=l$ is fixed. Similarly to that of (\ref{adt1}), we have
\begin{equation}
\frac{c_{n}}{n}\sum_{k=1}^{n}G( X_{nk})\, v_k\,\frac{1}{\sqrt{l_{n}}}\sum_{j=1}^{l_{n}}K\left[c_{n}(k/n-\tau_{j})\right]
K^{\ast}\left[c_{n}(k/n-\tau^{\ast})\right]=o_P(1).\label{double1}
\end{equation}

The proof of (\ref{adt1}) and (\ref{double1}) will be given in Section \ref {sect6.9}.
Equations (\ref{double2}) and (\ref{double1}) show the effect of
employing ``\textit{double trimming}''\ i.e. sample functionals
that involve two kernel functions, which will be used in the proofs of Theorems  \ref{Thm1} and \ref{Thm2}, (\ref {ad21}) and (\ref {ad22}).

\end{rem}

\subsection{Proof of Theorem \ref {lemma1}}

We start with the limit result for $S_{1n,l_{n}}$, i.e., (\ref {Lemma1a}).
 For $\mathbb{\alpha}\in\mathbb{R}^{p}$, let $v_k=\alpha'g(x_k)$. Since $\{x_k\}_{k\ge 1}$ is an ergodic stationary sequence with $E\|g(x_k)\|^{2+\delta}<\infty$ for some $\delta>0$, it is readily seen that  $\{v_k\}_{k\ge 1}$ is stationary  and ergodic, and condition
 (b) of Lemma \ref {thm27} holds with $A_0=Ev_1$ (see, for instance, Kallenberg (2002, Chapter 10)). (\ref {Lemma1a})  follows from Lemma \ref {thm27} with $G(x)\equiv 1$.

   We next consider $M_{1n,l_{n}}$, i.e., (\ref {Lemma1b}).
Set $Q_{k,n}:=\sqrt{\frac{c_{n}}{nl_n}}\mathbb{\alpha}^{\prime}g(x_{k})K_{kn}$
where $\mathbb{\alpha}\in%TCIMACRO{\U{211d}}%
%BeginExpansion
\mathbb{R}%EndExpansion
^{p}$. Note that
\begin{eqnarray}
\sum_{k=1}^{n}Q_{k,n}^{2}&=&\frac{c_{n}}{n}\sum_{k=1}^{n}\left[\mathbb{\alpha}^{\prime}g(x_{k})\right]^{2}
\frac{1}{l_{n}}\sum_{j=1}^{l_{n}}K^{2}\left[c_{n}(k/n-\tau_{j})\right]+o_{P}(1)
\nonumber\\
&=&E\left[\mathbb{\alpha}^{\prime}g(x_{k})\right]^{2}\int K^{2}+o_{P}(1) \la {97}
\end{eqnarray}
by using  Lemmas \ref {thm27} and \ref {thm27a} with $G(x)\equiv 1$,  $v_k=\left[\mathbb{\alpha}^{\prime}g(x_{k})\right]^{2}$ and $A_0=E\left[\mathbb{\alpha}^{\prime}g(x_{k})\right]^{2}$.   It follows from Hall
and Heyde (1980, Theorem 3.2) or Wang (2014, Theorem 2.1) that, equation\ (\ref{Lemma1b})
will follow, if we prove
\be
\max_{1\leq k\leq n}\left\vert Q_{k,n}\right\vert =o_{P}(1). \la {98}
\ee
 Note that for any $A>0$,
\begin{eqnarray*}
\max_{1\leq k\leq n}\left\vert Q_{k,n}\right\vert &\leq& \left\{ \sum_{k=1}^{n}Q_{k,n}^{2}I\left\{ \left\Vert g(x_{k})\right\Vert >A\right\} \right\} ^{1/2}+\left\{ \sum_{k=1}^{n}Q_{k,n}^{4}I\left\{ \left\Vert g(x_{k})\right\Vert \leq A\right\} \right\} ^{1/4}\\
&=:&II_{1n}(A)^{1/2}+II_{2n}(A)^{1/4}.
\end{eqnarray*}
Similar arguments used in (\ref {97}) show that the first term as $n\rightarrow\infty$ first
and then as $A\rightarrow\infty$
\begin{eqnarray*}
II_{1n}(A)&\leq& \|\alpha\|^2\frac{c_{n}}{n}\sum_{k=1}^{n}\left\Vert g(x_{k})\right\Vert ^{2}I\left\{ \left\Vert g(x_{k})\right\Vert >A\right\} \frac{1}{l_{n}}\sum_{j=1}^{l_{n}}K^2\left[c_{n}(k/n-\tau_{j})\right]+o_{P}(1)\\
&=& \|\alpha\|^2E\left\Vert g(x_{k})\right\Vert ^{2}I\left\{ \left\Vert g(x_{k})\right\Vert >A\right\} \int K^{2}+o_{P}(1)=o_{P}(1).
\end{eqnarray*}
By Lemma \ref {thm27a} with $G(x)\equiv 1$ and $v_k=1$, as $n\rightarrow\infty$, the second term
\[
II_{2n}(A)\leq  \|\alpha\|^4A^{4}\left(\frac{c_{n}}{nl_n}\right)^{2}\sum_{k=1}^{n}K_{kn}^{4}=o_{P}(1).
\]
Combining these facts together, we establish (\ref {98}). The proof of Theorem \ref {lemma1} is complete. $\Box$

\subsection{Proof of Theorem  \ref{lemma2}}

The result for $S_{2n,l_{n}}$, i.e., (\ref {Lemma2a}) follows from Lemma \ref {thm27} with $v_k\equiv 1$.

We next consider $M_{2n,l_{n}}$, i.e., (\ref {Lemma2b}). Set $Q_{k,n}:=\sqrt{\frac{c_{n}}{nl_n}}\mathbb{\alpha}^{\prime}g(X_{nk})K_{kn}$
where $\mathbb{\alpha}\in%TCIMACRO{\U{211d}}%
%BeginExpansion
\mathbb{R}%EndExpansion
^{p}$. Noting that $\int_0^1 g(X_{n, [nt]})dt$
is a continuous functional of $X_{n,\left[nt\right]}$,
the limit result of (\ref{Lemma2b}), jointly with  (\ref{Lemma2a}), will follow if we prove that
\begin{equation}
\Big\{ X_{n,\left[nt\right]},~\sum_{k=1}^{n}Q_{k,n}u_{k}\Big\} \Rightarrow\left\{ X_{t},~\mathbf{MN}\left(0,\sigma_u^2\int\left[\mathbb{\alpha}^{\prime}g(X_{t})\right]^{2}dt\int K^{2}\right)\right\} \label{ProofLemma2.10}
\end{equation}
on $D_{%
%TCIMACRO{\U{211d} }%
%BeginExpansion
\mathbb{R}
%EndExpansion
^{2}}[0,1]$.
First note that, by using Lemmas \ref {thm27} and \ref{thm27a} with $v_k\equiv 1$,
\begin{eqnarray*}
\sum_{k=1}^{n}Q_{k,n}^{2}&=&
\frac{c_{n}}{n}\sum_{k=1}^{n}\left[\mathbb{\alpha}^{\prime}
g(X_{nk})\right]^{2}\frac{1}{l_{n}}\sum_{j=1}^{l_{n}}K^{2}\left[c_{n}(k/n-\tau_{j})\right]+o_{P}(1)
\\
&=&
\int_0^1\left[\mathbb{\alpha}^{\prime}g(X_{n, [nt]})\right]^{2}dt\int K^{2}+o_{P}(1)
\rightarrow_{d} \int_0^1\left[\mathbb{\alpha}^{\prime}g(X_{t})\right]^{2}dt\int K^{2},
\end{eqnarray*}
indicating that
\bestar
 \Big\{ X_{n,\left[nt\right]},~\sum_{k=1}^{n}Q_{k,n}^{2}\Big\}  \Rightarrow\left\{ X_{t},\int_0^1\left[\mathbb{\alpha}^{\prime}g(X_{t})\right]^{2}dt\int K^{2}\right\}.
\eestar
 It follows from Theorem 2.1 of Wang
(2014), the limit result of\ (\ref{ProofLemma2.10})
will follow, if we prove
\begin{equation}
\max_{1\leq k\leq n}\left\vert Q_{k,n}\right\vert =o_{P}(1),\label{ProofLemma2.11}
\end{equation}
and
\begin{equation}
\frac{1}{\sqrt{n}}\sum_{k=1}^{n}\left\vert Q_{k,n}\right\vert =o_{P}(1).\label{ProofLemma2.12}
\end{equation}

In fact, by recalling the fact that $||g||^4$ is still continuous, it follows from Lemma \ref {thm27a} with $v_k=1$ again that
\[
\Big[\max_{1\leq k\leq n}\left\vert Q_{k,n}\right\vert \Big]^{4}\leq\sum_{k=1}^{n}Q_{k,n}^{4}\leq \|\alpha\|^4\Big(\frac{c_{n}}{nl_n}\Big)^{2}\sum_{k=1}^{n}\left\Vert g(X_{nk})\right\Vert ^{4}K_{kn}^4=o_{P}(1),
\]
yielding (\ref {ProofLemma2.11}). Similarly, by recalling  $l_n/c_n\to 0$, we have
\begin{eqnarray*}
\frac{1}{\sqrt{n}}\sum_{k=1}^{n}\left\vert Q_{k,n}\right\vert &\leq& \|\alpha\|\frac{1}{\sqrt{n}}\sqrt{\frac{c_{n}}{nl_n}}\sum_{k=1}^{n}\left\Vert g(X_{nk})\right\Vert K_{kn}
\\
&=&\|\alpha\|\sqrt{\frac{l_{n}}{c_{n}}}\frac{c_{n}}{nl_n}\sum_{k=1}^{n}\left\Vert g(X_{nk})\right\Vert K_{kn}=O_{P}\Big(\sqrt{\frac{l_{n}}{c_{n}}}\Big)=o_{P}\left(1\right),
\end{eqnarray*}
which shows (\ref{ProofLemma2.12}). The proof of Theorem \ref {lemma2} is complete.  $\Box$

\subsection{Proofs of Theorem  \ref{lemma3} and (\ref {ad21})-(\ref{ad22})}

To show Theorem \ref {lemma3}, we only prove  (\ref{Lemma3a}) since (\ref {Lemma3b}) is a direct consequence of (\ref {Lemma3a}) and Theorem~\ref {lemma2}.

Notice that, by the  condition (b), we may write
\begin{eqnarray*}
&&\sum_{k=1}^{n}\pi\left(d_{n}\right)^{-1}g(x_{k})K_{kn}
 \Big[\frac{c_{n}}{nl_{n}},\text{ }\sqrt{\frac{c_{n}}{nl_{n}}}u_{k}\Big]
\\
&=&\sum_{k=1}^{n}\left[H(X_{nk})+\pi\left(d_{n}\right)^{-1}R(d_{n},X_{nk})\right]K_{kn}
 \Big[\frac{c_{n}}{nl_{n}},\text{ }\sqrt{\frac{c_{n}}{nl_{n}}}u_{k}\Big] \\
&=&\sum_{k=1}^{n}H(X_{nk})K_{kn} \Big[\frac{c_{n}}{nl_{n}},\text{ }\sqrt{\frac{c_{n}}{nl_{n}}}u_{k}\Big]+(\Delta_{1n}, \Delta_{2n}),
\end{eqnarray*}
where $\ R(\lambda,x)=[R_{1}(\lambda,x),...,R_{p}(\lambda,x)]^{\prime}$ and
\bestar
\Delta_{1n} &=&\frac{c_{n}}{nl_{n}}\sum_{k=1}^{n}\pi\left(d_{n}\right)^{-1}R(d_{n},X_{nk})\,K_{kn}, \\
\Delta_{2n} &=&\sqrt {\frac{c_{n}}{nl_n}}\, \sum_{k=1}^{n}\pi\left(d_{n}\right)^{-1}R(d_{n},X_{nk})\,K_{kn}\, u_k.
\eestar
Now  (\ref{Lemma3a})  follows from Theorem \ref{lemma2} with $g(x)=H(x)$ if we prove
\be
|\al'\Delta_{in}| &=&o_P(1), \quad i=1,2, \la {adf1}
\ee
for any   $\al=(\al_1,...,\al_p)'\in %TCIMACRO{\U{211d}}%
%BeginExpansion
\mathbb{R}%EndExpansion
^{p}$.

We only prove (\ref {adf1}) with $i=2$ since the proof of $|\al'\Delta_{1n}|=o_P(1)$ is similar except simpler.
Recall $K_{kn}:=\sum_{j=1}^{l_{n}}K\left[c_{n}(k/n-\tau_{j})\right]$
and set, for $A>0$,
\[
\widetilde{R}_{n,l_{n}}(A)=\sqrt{\frac{c_{n}}{nl_n}}\sum_{k=1}^{n}\alpha^{\prime}\pi\left(d_{n}\right)^{-1}R(d_{n},X_{nk})I\left\{ \left\vert X_{nk}\right\vert \leq A\right\}K_{kn}u_{k}.
\]
Note that as $n\rightarrow\infty$ first and then $A\rightarrow\infty$
\begin{equation}
P\left(\al'\Delta_{2n}\neq\widetilde{R}_{n,l_{n}}(A)\right)\leq P\left(\max_{1\leq k\leq n}\left\vert X_{nk}\right\vert \geq A\right)\rightarrow0.\label{ProofLemma3.2}
\end{equation}
For any $\epsilon>0$\ and $A>0$, we have
\[
P\left(\left\vert\al'\Delta_{2n}\right\vert \geq\epsilon\right)\leq P\left(\al'\Delta_n\neq\widetilde{R}_{n,l_{n}}(A)\right)+\epsilon^{-2}E\left[\widetilde{R}_{n,l_{n}}(A)\right]^{2}.
\]
Now $|\al'\Delta_{2
n}| =o_P(1)$ follows from (\ref{ProofLemma3.2})
and the fact that as $n\rightarrow\infty$ for any $A>0$
\begin{eqnarray*}
 E\left[\widetilde{R}_{n,l_{n}}(A)\right]^{2}
&\leq&\frac{c_{n}}{nl_n}C\sum_{k=1}^{n}E\left|\alpha^{\prime}\pi\left(d_{n}\right)^{-1}R(d_{n},X_{nk})\right| ^{2}I\left\{ \left\vert X_{nk}\right\vert \leq A\right\} K_{kn}^{2}
\\
&\leq&\frac{c_{n}}{n}C\|\alpha\|^2\left(1+A^{\delta}\right)^{2}\epsilon_{n}^{2}\frac{1}{l_n}\sum_{k=1}^{n}K_{kn}^{2}\rightarrow0,
\end{eqnarray*}
where $\epsilon_{n}=\max_{1\le i\le p} |[\pi_i(d_n)]^{-1}a_i(d_n)|\rightarrow 0$ and
 we have used (\ref{mai9a1}) of Lemma \ref {thm27a} (with $G(x)\equiv 1$ and $v_k\equiv 1$).
The proof of Theorem \ref {lemma3} is now complete.

\bigskip
{\it Proofs of (\ref {ad21}) and (\ref {ad22})}  are essentially the same as that of (\ref{Lemma3a}).
We only provide a outline for (\ref {ad21}).
For any $\al, \beta\in \mathbb{R}$, let
\bestar
\widetilde Q_{k,n} &=& \sqrt{\frac{c_n}{nl_n}} \Big(\al H_2(X_{n,k})K_{kn}+\beta\,
K^*_{kn}\Big),
\eestar
where $K_{kn}^{\ast}:=\sum_{j=1}^{l_{n}}K^{\ast}\left[c_{n}\left(k/n-\tau_{j}\right)\right]$.
As in the proof of (\ref{Lemma3a}), we have
\bestar
\al\, U_{1n}+\beta\, U_{2n} &=&  \sum_{k=1}^n \widetilde Q_{k,n}\, u_k +o_P(1).
\eestar
Note that, by using (\ref {double2}) and Lemmas \ref {thm27} and \ref {thm27a},
\bestar
 \sum_{k=1}^n \widetilde Q_{k,n}^2 &=& \al^2\, \int_0^1H_2^2(X_{n, [nt]})dt\, \int K^2+ 2\al \beta\,
 \int_0^1H_2(X_{n, [nt]})dt\, \int KK^*\no\\
 &&\qquad \qquad+\beta^2 \int (K^*)^2+o_P(1),
\eestar
indicating
\bestar
 \Big\{ X_{n,\left[nt\right]},~\sum_{k=1}^{n}\widetilde Q_{k,n}^{2}\Big\}  \Rightarrow\left\{ X_{t},~[\al, \beta]\, V_1\,[\al, \beta]'\right\}.
\eestar
Similarly, we may prove that (\ref {ProofLemma2.11}) and (\ref {ProofLemma2.12}) hold with $Q_{kn}$ being replaced by $\widetilde Q_{k,n}$.
As a consequence, (\ref {ad21}) follows from Wang (2014) as in the proof of  Theorem \ref {lemma2}.
$\Box$

\subsection{Proofs of Theorems \ref{Thm1} and \ref{Thm2}}

We only prove Theorem \ref{Thm2} since the proof of Theorem \ref {Thm1} is similar except simpler. Let
\bestar
A_{1n} &=& \frac {c_n}{nl_n} \sum_{k=1}^{n}\pi(d_{n})^{-2}f^2(x_k)\, K_{kn}, \quad A_{2n} = \frac {c_n}{nl_n} \sum_{k=1}^{n}\pi(d_{n})^{-1}f(x_k)\, K_{kn}, \\
A_{3n} &=&  \frac {c_n}{nl_n^*} \sum_{k=1}^{n}\pi(d_{n})^{-1}f(x_k)\, K_{kn}^*, \\
B_{1n} &=& \sqrt{\frac{c_n}{nl_n}}\sum_{k=1}^{n}\pi(d_{n})^{-1}f(x_k)\,K_{kn}u_{k}, \quad B_{2n} = \sqrt{\frac{c_n}{nl_n^*}}\sum_{k=1}^{n}K_{kn}^{\ast}u_{k}.
\eestar
Recall (\ref {demeaning}) and $Z_{kn}=f(x_k)K_{kn}$ and note that $\frac {c_n}{nl_n^*}{\sum_{k=1}^{n}K_{kn}^{\ast}}=\int K^*+o(1)$. It is readily seen from (\ref{Lemma3a}) of Theorem \ref {lemma3} and Theorem \ref {lemma2} that
\be
&&\frac{1}{\lambda_{n}\pi^{2}(d_{n})}
\sum_{k=1}^{n}Z_{kn}\overline{f}_{k} \no\\
&=& \frac {c_n}{nl_n} \sum_{k=1}^{n}\pi(d_{n})^{-1}f(x_k)\, K_{kn}\,
\Big[\pi(d_{n})^{-1}f(x_k)-\frac{\sum_{k=1}^{n}\pi(d_{n})^{-1}f(x_k)K_{kn}^{\ast}}
{\sum_{k=1}^{n}K_{kn}^{\ast}}\Big] \no\\
&=& A_{1n}- A_{2n} A_{3n}\Big/  \int K^*+o_P(1)\no \\
&=& C_n\, \int K +o_P(1) ,  \la {ad40}
\ee
where
\bestar
C_n=\left\{ \begin{array}{ll}
 \int_{0}^{1}H^{2}(X_{n, [nt]})dt-\left[\int_{0}^{1}H(X_{n, [nt]})dt\right]^{2}, &\text{ if }K_{kn}^{\ast}=\sum_{j=1}^{l_{n}}K^{\ast}\left[c_{n}\left(k/n-\tau_{j}\right)\right],\\
 \int_{0}^{1}H^{2}(X_{n, [nt]})dt-\big[\int_{0}^{1}H(X_{n, [nt]})dt\big]\, H(X_{n,[n\tau^*]}), &\text{ if }K_{kn}^{\ast}=K^{\ast}\left[c_{n}\left(k/n-\tau^{\ast}\right)\right].
\end{array}\right.
\eestar
Similarly, we have
\be
&&\frac{\sqrt{\lambda_{n}^{\ast}}}{\lambda_{n}\pi(d_{n})}\sum_{k=1}^{n}Z_{kn}\overline{u}_{k} \no\\
&=& \frac{\sqrt{\lambda_{n}^{\ast}}}{\lambda_{n}}\left\{ \sum_{k=1}^{n}\pi(d_{n})^{-1}f(x_k) K_{kn}u_{k}-\frac{\left[\sum_{k=1}^{n}\pi(d_{n})^{-1}f(x_k)K_{kn}\right]\,\left[\sum_{k=1}^{n}K_{kn}^{\ast}u_{k}\right]}{\sum_{k=1}^{n}K_{kn}^{\ast}}\right\} \\
&=&\sqrt {\frac{l_{n}^{\ast}}{l_{n}}} \, B_{1n}-A_{2n}\, B_{2n}\big/  \int K^*+o_P(1) \no\\
&=& A_n\, B_n +o_P(1), \la {ad41}
\ee
where
\bestar
A_n &=& \left[R^{\ast},\text{ }-\frac{\int_{0}^{1}H\left(X_{n, [nt]}\right)dt\int K}{\int K^{\ast}}\right],~~~~
B_n =[B_{1n}, B_{2n}]'.
%\left[\begin{array}{l}
%B_{1n}\\
%B_{2n}
%\end{array}\right]=
%\left[\begin{array}{l}
%\sqrt{\frac{c_n}{nl_n}}\,\sum_{k=1}^{n}H(X_{nk})K_{kn}u_{k}\\
%\sqrt{\frac{c_n}{nl_n^*}}\,\sum_{k=1}^{n}K_{kn}^{\ast}u_{k}
%\end{array}\right].
\eestar
Since  both $C_n$ and $A_n$ are continuous functionals of $X_{n, [nt]}$, a simple application of  (\ref {ad21}) and (\ref {ad22}) yields that
\be
\sqrt{\lambda_{n}^{\ast}}\pi(d_{n})\left(\hat{\beta}-\beta\right) &=&
\frac{\frac{\sqrt{\lambda_{n}^{\ast}}}{\lambda_{n}\pi(d_{n})}\sum_{k=1}^{n}Z_{kn}\overline{u}_{k}}{\frac{1}{\lambda_{n}\pi(d_{n})^{2}}
\sum_{k=1}^{n}Z_{kn}\overline{f}_{k}} \no\\
&=& \Big( C_n\int K\Big)^{-1}\, A_n\, B_n +o_P(1) \no\\
&\to_d& \sigma_{u}\, \mathbf{MN}\left(\mathbf{0},~\Big(C\int K\Big)^{-2}\,AVA'\right), \la {ad42}
\ee
as required. The proof of Theorem \ref {Thm2} is complete.
$\Box$

\subsection{Proof of Theorem \ref{thm3}}

We only prove Theorem \ref{thm3} under conditions of Theorem \ref{Thm2} since the proof  under conditions of Theorem \ref{Thm1} is similar.
In addition to $A_{2n}, B_{1n}$, $B_{2n}$, $A_n$ and $B_n$ in the proof of Theorem \ref {Thm2}, we  define
\bestar
V_n &=&  \begin{bmatrix}  %\left[\begin{array}{cc}
\int_{0}^{1}H^{2}\left(X_{n, [nt]}\right)dt\int K^{2} & \int_{0}^{1}H\left(X_{n, [nt]}\right)dt\, Q^*\\
\int_{0}^{1}H\left(X_{n, [nt]}\right)dt\, Q^* & \int\left(K^{\ast}\right)^{2}
 \end{bmatrix}.
\eestar
As in the proof of (\ref {ad40}),  by letting $D_{n}=$diag$\left\{ \pi(d_{n})\sqrt{\lambda_{n}},\sqrt{\lambda_{n}^{\ast}}\right\} $, we have
\be
&&\frac{\lambda_{n}^{\ast}}{\lambda_{n}^{2}\pi^2(d_{n})}\mathcal{A}_{n}\mathcal{V}_{n}\mathcal{A}_{n}^{\prime}
\ =\ \frac{\lambda_{n}^{\ast}}{\lambda_{n}^{2}
\pi(d_{n})^{2}}\mathcal{A}_{n}D_{n}D_{n}^{-1}\mathcal{V}_{n}D_{n}^{-1}D_{n}\mathcal{A}_{n}^{\prime} \no\\
&=&\left[\sqrt{\frac {\lambda_{n}^*}{\lam_n}},\text{ }-\frac{\frac {1}{\lam_n \pi(d_n) }\, \sum_{k=1}^{n}f(x_{k})K_{kn}}{\frac {1}{\lam_n^*}\,
\sum_{k=1}^{n}K_{kn}^{\ast}}\right]
\nonumber\\
&&~~~~~~~~\times\left[\begin{array}{ll}
\frac{1}{\lambda_{n}\pi^2(d_n)}\sum_{k=1}^{n}K_{kn}^{2}f^2\left(x_{k}\right) & \frac{1}{\pi(d_n)\sqrt{\lambda_{n}\lambda_{n}^{\ast}}}\sum_{k=1}^{n}K_{kn}^{\ast}K_{kn}f\left(x_{k}\right)\\
\frac{1}{\pi(d_n)\sqrt{\lambda_{n}\lambda_{n}^{\ast}}}\sum_{k=1}^{n}K_{kn}^{\ast}K_{kn}f\left(x_{k}\right) & \frac{1}{\lambda_{n}^{\ast}}\sum_{k=1}^{n}\left(K_{kn}^{\ast}\right)^{2}
\end{array}\right]
\nonumber\\
&&~~~~~~~~\times\left[\sqrt{\frac {\lambda_{n}^*}{\lam_n}},\text{ }-\frac{\frac {1}{\lam_n \pi(d_n) }\, \sum_{k=1}^{n}f(x_{k})K_{kn}}{\frac {1}{\lam_n^*}\,
\sum_{k=1}^{n}K_{kn}^{\ast}}\right]
\nonumber\\
&=& A_n \, V_n\, A_n'+o_P(1). \la {ad31}
\ee
Since $\tilde{\sigma}^{2}=\sigma_u^{2}+o_{P}(1)$ under given assumptions, by using  the similar arguments as in the proofs of (\ref {ad41}) and (\ref {ad42}), it follows from (\ref {ad31}) that
\bestar
\hat{T}=
\frac{\frac{\sqrt{\lambda_{n}^{\ast}}}{\lambda_{n}\pi(d_{n})}\sum_{k=1}^{n}Z_{kn}\overline{u}_{k}}
{\sqrt {\tilde{\sigma}^{2}\,\frac{\lambda_{n}^{\ast}}{\lambda_{n}^{2}\pi^2(d_{n})}\mathcal{A}_{n}\mathcal{V}_{n}\mathcal{A}_{n}^{\prime}}}
= \big(\sigma_u^2\, A_n \, V_n\, A_n'\big)^{-1/2}\, A_n\, B_n+o_P(1)
\to_d N(0,1),
\eestar
as required.
$\Box$

\subsection{Proof of Lemma \ref {thm27}}

%Throughout the proofs, we make use of the notations: if $a,b$ are not integers, $\sum\limits_{k=a}^{b}=\sum\limits_{k=[a]+1}^{[b]}$;
%where $[a]$ and $[b]$ denote the integer part of $a$ and $b$ respectively.

 We only prove (\ref {mai1}), as the proof of (\ref {mai}) is similar except more simpler.
 We start with the proof of (\ref {mai1}) by  assuming that there exists an $A>0$ such that $K(x)=0$ if $|x|\ge A$ and $K(x)$ is Lipschitz continuous on $%BeginExpansion
\mathbb{R}%EndExpansion
$. This restriction will be removed later.

  Without loss of generality,  suppose  $A=1$.  Set   $\de_{1n, j}=[n(\tau_j-1/c_n)]\vee 1$, $\de_{2n, j}=[n(\tau_j+1/c_n)]\vee 1$ and $\de_{n, j}=[n\tau_j]$. Recall $\tau_j=j/(l_n+1)$.   Since
\be
|c_n(k/n-\tau_j)|< 1 \quad \mbox{only if}\quad \de_{1n, j}\le  k\le  \de_{2n, j}, \quad j=1, ..., l_n, \label{domaink}
\ee
by letting $R_{1n, j}=\frac {c_n}n \sum_{k=\de_{1n, j}}^{\d_{2n,j}}\ v_k\ K\big[c_n(k/n-\tau_j)\big]$ and
\bestar
R_{2n, j}&=&\frac {c_n}n \sum_{k=\de_{1n, j}}^{\d_{2n, j}} \big[G\big( X_{nk}\big)- G\big( X_{n, \de_{n, j}}\big)\big]\ v_k\ K\big[c_n(k/n-\tau_j)\big],
\eestar
we have
\bestar
S_{n, l_n} &=& \frac{1}{l_{n}}\sum_{j=1}^{l_{n}}\, \frac {c_n}n  \sum_{k=1}^n G( X_{nk})\, v_k\,K\big[c_n (k/n-\tau_j)\big]\no\\
&=& \frac{1}{l_{n}}\sum_{j=1}^{l_{n}}\,  G\big( X_{n, \de_{n,j}}\big)\frac {c_n}n \sum_{k=\de_{1n, j}}^{\d_{2n,j}}\ v_k\ K\big[c_n(k/n-\tau_j)\big] \no\\
&&\qquad + \frac{1}{l_{n}}\sum_{j=1}^{l_{n}}\,
\frac {c_n}n \sum_{k=\de_{1n, j}}^{\d_{2n, j}} \big[G\big( X_{nk}\big)- G\big( X_{n, \de_{n, j}}\big)\big]\ v_k\ K\big[c_n(k/n-\tau_j)\big]\no\\
&=&\frac{1}{l_{n}}\sum_{j=1}^{l_{n}} G\big( X_{n, \de_{n,j}}\big)\, R_{1n, j} + \frac{1}{l_{n}}\sum_{j=1}^{l_{n}}\, R_{2n, j} \no\\
&=&  \frac{1}{l_{n}}\sum_{j=1}^{l_{n}}\,  G\big( X_{n, \de_{n,j}}\big)\, A_0\int K\,  +
\frac{1}{l_{n}}\sum_{j=1}^{l_{n}} G\big( X_{n, \de_{n,j}}\big)\, \big[R_{1n, j} -A_0\int K\big] + \frac{1}{l_{n}}\sum_{j=1}^{l_{n}}\, R_{2n, j} \no\\
&:=&  \frac{1}{l_{n}}\sum_{j=1}^{l_{n}}\,  G\big( X_{n, \de_{n,j}}\big)\, A_0\int K\,  +R_{1n} +R_{2n}.
\eestar
Since   $ \frac{1}{l_{n}}\sum_{j=1}^{l_{n}}\,  G\big( X_{n, \de_{n,j}}\big) =\int_0^1 G(X_{n, [nt]})dt+o_P(1)\to_d \int_0^1G(X_t)dt$, it suffices to show that
\be
R_{jn} &=&o_P(1), \quad j=1,\ 2. \la{ad1}
\ee

To prove (\ref {ad1}), we start with some preliminaries.    Recalling $X_{n,[nt]}\Rightarrow X_t$ on $D_{%
%TCIMACRO{\U{211d} }%
%BeginExpansion
\mathbb{R}
%EndExpansion
^{p}}[0,1]$ and  the limit process $X(t)$ is path continuous,   we have
$X_{n,[nt]}\Rightarrow X_t$ on $D_{%
%TCIMACRO{\U{211d} }%
%BeginExpansion
\mathbb{R}
%EndExpansion
^{p}}[0,1]$ in the sense of uniform topology. See, for instance,  Section 18 of Billingsley (1968).  This fact implies  that
\be
\limsup_{N\rightarrow\infty}\limsup_{n\rightarrow\infty}P\Big(\max\limits_{1\le k\le n} ||X_{nk}||\ge N\Big)=0,
%\limsup_{n\rightarrow\infty}P\Big(\max\limits_{1\le k\le n} ||X_{nk}||\in  E\Big)=0,
\label{leK}
\ee
%where $E$ is an arbitrary  subset of ${\mathbb{R}}$ with Lebesgue  measure zero,
 and by the tightness of
$\{X_{n,[nt]}\}_{ 0\le t\le 1}$, for any $\varepsilon>0$ and $\delta>0$, there is some $\tilde{\delta}=\tilde{\delta}(\varepsilon,\delta)>0$ such that
\be
P(\sup_{|s-t|\le \tilde{\delta}} ||X_{n,[nt]}-X_{n,[ns]}|| \ge \delta)\le \varepsilon   \la {leK1}
\ee
holds for all sufficiently large $n$. In terms of   (\ref {leK1}),  for any $\delta>0$,  we have
\be
\lim_{n\to\infty }P(\max_{1\le j\le l_n} \max_{\de_{1n, j}\le l\le  k\le \de_{2n, j}}  ||X_{nk}-X_{nl}||\ge \delta)= 0. \label{atightness2}
\ee

We are now ready to prove (\ref {ad1}), starting with $j=1$.

For any $N>0$, we let  $G_N(x)=G(x)\xi_N(x)$ with
\bestar
\xi_N(x)=\left\{
\begin{array}{ll}
1, & ||x||\le N,\\
2-||x||/N, & N<||x||<2N,\\
0, & ||x||\ge 2N,
\end{array}
\right.
\eestar
 and
\bestar
\widetilde R_{1n}&=& \frac{1}{l_{n}}\sum_{j=1}^{l_{n}} G_N\big( X_{n, \de_{n,j}}\big)\, \big[R_{1n, j} -A_0\int K\big].
\eestar
Note that, as $n\to\infty$ first and then $N\to\infty$,
\be
P(R_{1n}\not = \widetilde R_{1n}) \le P\Big(\max\limits_{1\le k\le n} ||X_{nk}||\ge N\Big) \to 0, \la {ad2}
\ee
 and
\be
|\widetilde R_{1n}|&\le & \frac{C_N}{l_{n}}\sum_{j=1}^{l_{n}} \big| R_{1n, j} -A_0\int K\big|, \la {ad3}
\ee
where $C_N:=\sup_{x} |G_N(x)|<\infty$ is a constant depending only on $N$,  due to the continuity of $G(x)$.  Result (\ref {ad1}) with $j=1$ will follow if we prove
\be
\max_{1\le j\le l_n} E \big| R_{1n, j} -A_0\int K\big| &\to&  0, \la {ad4}
\ee
as $n\to\infty$. Indeed, by virtue of (\ref {ad3}) and (\ref {ad4}), we have $E|\widetilde R_{1n}|\to 0$ and then $\widetilde R_{1n}=o_P(1)$ for each $N\ge 1$. This, together with (\ref {ad2}), yields $ R_{1n}=o_P(1)$.

Since, as $n\to\infty$,
\be
\max_{1\le j\le l_n} \Big| \frac {c_n}n \sum_{k=\de_{1n, j}}^{\d_{2n, j}}\ K\big[c_n(k/n-\tau_j)\big] -\int K \Big| \to 0, \label{Rie}
\ee
to prove (\ref {ad4}), it suffices to show that $\max_{1\le j\le l_n} E|A_n(\tau_j)| \to 0$, where
\bestar
A_n(\tau_j) &=& \frac {c_n}n \sum_{k=\de_{1n, j}}^{\d_{2n, j}}\ (v_k-A_0)\ K\big[c_n(k/n-\tau_j)\big].
\eestar

Let $\ga=\ga_n$ be integers such that $ \ga \to \infty$ and $\ga\,c_n/n\to 0$,   $T_{1n, j}=[\de_{1n, j}/\ga]$ and $T_{2n, j}=[\de_{2n, j}/\ga]$. Noting (\ref{domaink}),  we  may write
\bestar
A_{n} (\tau_j)&=&\frac {c_n}n \sum_{k=\de_{1n, j}}^{\de_{2n, j}} (v_k-A_0)\,  K\big[c_n(k/n-\tau_j)\big] \no\\
&=&\frac {c_n}n \sum_{s=T_{1n,j }}^{T_{2n, j}}\sum_{k=s\ga}^{(s+1)\ga} (v_k-A_0)\, K\big[c_n(k/n-\tau_j)\big]% +\frac {c_n}n\sum_{k=\ga T_{2n,j}}^{\de_{2n, j}}  (v_k-A_0)\,  K\big[c_n(k/n-\tau_j)\big]
 \no\\
&\le&\frac {\ga c_n}n \sum_{s=T_{1n,j}}^{T_{2n,j}}K\big[c_n(s\ga/n-\tau_j)\big] \frac 1{\ga} \Big|\sum_{k=s\ga}^{(s+1)\ga}(v_k-A_0)\, \Big| \no\\
&&+  \frac {c_n}n \sum_{s=T_{1n,j}}^{T_{2n,j}}\sum_{k=s\ga}^{(s+1)\ga}  |v_k-A_0|\, \Big| K\big[c_n(k/n-\tau_j)\big] -K\big[c_n(s\ga/n-\tau_j)\big]\Big| \no\\
&:=&A_{1n} (\tau_j)+A_{2n}(\tau_j).
\eestar
Recall $\sup_{k\ge 1}E|v_k|<\infty$ by condition (b), it is readily from the
 the Lipschitz condition on $K(x)$ that
\bestar
&&E A_{2n}(\tau_j)
\le C\, \frac {\ga c_n}n\, \frac {c_n}n \sum_{k=\de_{1n, j}}^{\d_{2n,j}} E |v_k-A_0| %+
% \frac {c_n}n\sum_{k=\ga T_{2n,j}}^{\de_{2n, j}}E|v_k-A_0|\,  K\big[c_n(k/n-\tau_j)\big]  \no\\
\le C\, \frac {\ga c_n}n \to 0,
\eestar
uniformly in $1\le j\le l_n$. Similarly, by using  condition (b),  we have
\bestar
\max_{1\le j\le l_n}EA_{1n} (\tau_j)  &\le &  \max_{\ga\le s\le n-\ga}  E \, \big|  \frac 1\ga  \sum_{k=s}^{s+\ga} v_k-A_0\big| \,\max_{1\le j\le l_n} A_{4n} (\tau_j)   \to 0,
\eestar
where
\bestar
A_{4n} (\tau_j) &=& \frac {\ga c_n}n  \sum_{s=T_{1n, j}}^{T_{2n, j}}K\big[c_n(s\ga/n-\tau_j)\big],
\eestar
and we have used the fact that %uniformly in $1\le j\le l_n$,
%\bestar
%A_{4n} (\tau_j) &\le & \frac {\ga\, c_n}n  \int_{T_{1n,j}/n}^{T_{2n,j}/n}K\big[c_n(x\ga-\tau_j)\big] dx+O(1) \no\\
%&\le& \frac {c_n}n  \int_{\ga T_{1n,j}/n}^{\ga T_{2n,j}/n}K\big[c_n(x-\tau_j)\big] dx+O(1) \no\\
%&\le& \frac {c_n}n  \int_{\tau_j-1/c_n}^{\tau_j+1/c_n}K\big[c_n(x-\tau_j)\big] dx+O(1) =O(1).
%\eestar
$
\max_{1\le j\le l_n} \Big|A_{4n} (\tau_j) -\int K \Big| \to 0.
$
Combining all these facts, we prove (\ref {ad4}), and complete the proof of $ R_{1n}=o_P(1)$.

We next show $R_{2n}=o_P(1)$.   Let
 $ \widetilde R_{2n}=\frac{1}{l_{n}}\sum_{j=1}^{l_{n}}\, \widetilde R_{2n, j}$, where
\bestar
\widetilde R_{2n, j} =\frac {c_n}n \sum_{k=\de_{1n, j}}^{\d_{2n, j}} \big[G_N\big( X_{nk}\big)- G_N\big( X_{n, \de_{n,j}}\big)\big]\ v_k\ K\big[c_n(k/n-\tau_j)\big].
\eestar
In terms of  (\ref {leK}), we have
\bestar
P(R_{2n}\not = \widetilde R_{2n}) \le P\Big(\max\limits_{1\le k\le n} ||X_{nk}||\ge N\Big) \to 0,
\eestar
as $n\to\infty$ first and then $N\to\infty$.  Result  $R_{2n}=o_P(1)$  will follow if we prove
$
\widetilde R_{2n}=o_P(1),$
for each fixed $N\ge 1$.

Recall  that $G_N(x)$ is continuous with compact support.  For any $\ep>0$, % there exists a continuous function ${G}_{N,\ep}(x)$   such that
%$
%|G_N(x)-{G}_{N,\ep}(x)|\le \ep
%$ except in a Lebeque measure zero set $E$.
%Since ${G}_{N,\ep}(x)$ is uniformly continuous with regard to $x$, we further have that, for any $\ep>0$,
 there exists a $\de_\ep>0$ so that $|G_N(x)-G_N(y)|\le \ep$ whenever $||x-y||\le \de_\ep$.
Write
\bestar
\Omega_{\delta_\ep} &=& \{\omega:  \max_{1\le j\le l_n}\max_{\de_{1n, j}\le l\le  k\le \de_{2n,j}}  ||X_{nk}-X_{nl}||\le  \delta_{\ep}\}.
\eestar  By virtue of the facts above and (\ref{Rie}), it is readily seen that
\bestar
&&\max_{1\le j\le l_n}E\big[| \widetilde R_{2n,j }|I(\Omega_{\delta_\ep})\big]\no\\
&\le & E\Big\{
\max_{1\le j\le l_n}\max_{\de_{1n,j}\le l\le  k\le \de_{2n, j}}|G_N(X_{nk})-G_N(X_{nl})| \, \frac {c_n}n \sum_{k=\de_{1n, j}+1}^{\d_{2n, j}} |v_k|\,   K\big[c_n(k/n-\tau_j)\big]\Big\}\no\\  &\le& {\ep} \,\sup_{k\ge 1}E|v_k|\,\frac {c_n}n \sum_{k=\de_{1n, j}+1}^{\d_{2n, j}}  K\big[c_n(k/n-\tau_j)\big] \le C_N\ep ,
\eestar
where $C_N$ is a constant depending only on $N$.
Now,  for any $\eta_1>0$ and $ \eta_2>0$, let $\ep=\eta_1\eta_2$ and $n_0$ be large enough so that, for all $n\ge n_0$ [recall (\ref {atightness2})],
 \bestar
 P\big( \max_{1\le j\le l_n}\max_{\de_{1n,j}\le l\le  k\le \de_{2n, j}}  ||X_{nk}-X_{nl}||\ge \delta_\ep\big)\le \eta_2.
 \eestar
 It is readily seen that, for all $n\ge n_0$,
\bestar
P(|\widetilde R_{2n}|\ge \eta_1) &\le& P\big (\bar  \Omega_{\delta_\ep}\big) +\eta_1^{-1}\, \frac{1}{l_{n}}\sum_{j=1}^{l_{n}} E\big[| \widetilde R_{2n,j }|I(\Omega_{\delta_\ep})\big]  \ \le\ C_{N}\, \eta_2
\eestar
 where $\bar  \Omega_{\delta_\ep}$ denotes the complementary set of $  \Omega_{\delta_\ep}$ and $C_{N}$ is a constant depending only on $N$. This yields $
\widetilde R_{2n}=o_P(1),$
for each fixed $N\ge 1$, and completes the proof of $R_{2n}=o_P(1)$ .

   We finally remove the restriction on $K$ and then conclude the proof of Lemma \ref {thm27}.
If $K$ has compact support, then there exists  $A_1>0$ such that $K(x)=0$ holds for all $|x|\ge A_1$.
If $K$ is eventually monotonic,  then for any $\ep>0$,  we can also choose a constant $A_1:=A_1(\ep)>0$ such that $K(x)$ is monotonic on
$(-\infty, -A_1)$ and $(A_1,\infty)$ and $\int_{|x|>A_1} K(x)dx<\ep$ (in order to simplify the notations, here we use the same notation $A_1$ to denote the constant).

 Since $K\ge 0$ with $\int K<\infty$,  for any $\ep>0$, there exists an $A:=A_\ep\ge A_1+1$ such that
\bestar
\int |K-K_{\ep, A}| \le \ep,
\eestar
where $K_{\ep, A} (x)=0$  if $|x|\ge A$ and $K_{\ep, A}(x)$ is Lipschitz continuous on
 ${%
%TCIMACRO{\U{211d} }%
%BeginExpansion
\mathbb{R}
%EndExpansion
}$. Let $\widetilde K(x)=K(x)-K_{\ep, A} (x) $ and
\bestar
S_{n, \ep} &=& \frac{1}{l_{n}}\sum_{j=1}^{l_{n}}\, \frac {c_n}n  \sum_{k=1}^n G( X_{nk})\, v_k\,  \widetilde K\big[c_n (k/n-\tau_j)\big].
\eestar
It suffices to show that, as $n\to\infty$ first and then $\ep\to 0$,
\be
 S_{n, \ep} &=& o_P(1). \la {sa2}
\ee

The proof of  (\ref {sa2}) is similar to that of (\ref {ad1}). Indeed,  by letting% $G_N(x)=G(x)I(||x||\le N)$ and
\bestar
 S_{n, \ep, N} = \frac{1}{l_{n}}\sum_{j=1}^{l_{n}}\, \frac {c_n}n  \sum_{k=1}^n G_N( X_{nk})\, v_k\,  \widetilde K\big[c_n (k/n-\tau_j)\big],
\eestar
we have
\bestar
P\Big[ S_{n, \ep}\not = S_{n, \ep, N}\Big] \le P\Big(\max\limits_{1\le k\le n} ||X_{nk}||\ge N\Big) \to 0,
\eestar
 as $n\to\infty$ first and then $N\to\infty$. Hence it suffices to show that, for each fixed $N\ge 1$, $ S_{n, \ep, N}=o_P(1)$ as $n\to\infty$ first and then $\ep \to 0$.
Note that
 \bestar
 \sup_{1\le j\le l_n}\Big|\frac {c_n}n \sum_{k=1}^{n}\,  \big|\widetilde K\big[c_n (k/n-\tau_j)\big]\big|I(c_n |k/n-\tau_j|\le A)-\int_{-A}^A  |\widetilde K(x)|dx \Big| \rightarrow 0,
 \eestar
 and, if $K(x)$ is monotonic on $(-\infty, -A)$ and $(A,\infty)$ then for sufficiently large $n$, uniformly for $1\le j\le l_n$,
 \bestar
 &&\frac {c_n}n \sum_{k=1}^{n}\,\big|\widetilde K\big[c_n (k/n-\tau_j)\big]\big|I(c_n |k/n-\tau_j|>A)\\
 &=&\frac {c_n}n \sum_{k=1}^{n}\, K\big[c_n (k/n-\tau_j)\big]I(c_n |k/n-\tau_j|>A)\\
 &\le& \int_{|x|>A-c_n/n} K(x)dx \le \int_{|x|>A_1} K(x)dx<\ep.
 \eestar
Hence, in terms of the uniformed  boundedness of $G_N(x)$, we have
\bestar
E S_{n, \ep, N} \le C_N \, \sup_k E |v_k|\,\frac{1}{l_{n}}\sum_{j=1}^{l_{n}}\, \frac {c_n}n \sum_{k=1}^{n}\, \big|\widetilde K\big[c_n (k/n-\tau_j)\big]\big|  \to 0,
\eestar
as $n\to\infty$ first and then $\ep\to0$. Hence  $ S_{n, \ep, N}=o_P(1)$ as $n\to\infty$ first and then $\ep \to 0$.
The proof of  (\ref {sa2}) is completed.
$\Box$

\subsection{Proof of Lemma  \ref{thm27a}}
 We first prove (\ref {mai9a}). Using similar arguments as in the proof of   (\ref {ad1}) or (\ref {sa2}),  it suffices to show that, as $n\to\infty,$
\bestar
I_n &:=&\frac {c_n}n  \sum_{k=1}^n  \frac{1}{l_n}\sum_{1\le i<j\le l_n}K\big[c_n (k/n-\tau_i)\big]K\big[c_n (k/n-\tau_j)\big] \to  0.
\eestar
Take $\eta_{n,i,j}=\frac 12 n(\tau_i+\tau_j).$ Note that $c_n (k/n-\tau_i)\ge c_n(j-i)/(2l_n)$ if $k\ge \eta_{n,i,j}$ and $|c_n (k/n-\tau_j)|\ge c_n(j-i)/(2l_n)$ if $k\le \eta_{n,i,j}$. It follows from $K(x)\le C/(1+|x|)$ that
\bestar
I_n &=&   \frac{1}{l_n}\sum_{1\le i<j\le l_n} \frac {c_n}n \sum_{k=1}^n K\big[c_n (k/n-\tau_i)\big]K\big[c_n (k/n-\tau_j)\big] \no\\
&\le& \, \frac{C}{l_n}\sum_{1\le i<j\le l_n} \frac {l_n}{c_n(j-i)}\, \frac {c_n}n \sum_{k=1}^n\big( K\big[c_n (k/n-\tau_i)\big]+K\big[c_n (k/n-\tau_j)\big]\big)\no\\
&\le &\frac{C}{c_n}\,\sum_{1\le i<j\le l_n}  \frac 1{j-i} \le C\,  l_n\log l_n/c_n \to 0,
\eestar
as required.

The proof of (\ref {mai9}) is similar to that of (\ref {mai9a}) and hence the details are omitted.  Result (\ref {mai9a1}) follows easily from (\ref {mai9}) and   (\ref {mai9a}). As for  (\ref {mai9b}), it follows from the similar arguments as in the proof of (\ref {ad1})  and the fact: as $n\to\infty$,
\bestar
&& \big( \frac {c_n}n\big)^2  \sum_{k=1}^n  \,\Big(\frac{1}{\sqrt {l_n}}\sum_{j=1}^{l_n}K\big[c_n (k/n-\tau_j)\big]\Big)^4 \no\\
&\le& 2\, \big( \frac {c_n}n\big)^2  \sum_{k=1}^n  \Big(\frac{1}{\sqrt {l_n}}\sum_{j=1}^{l_n}K^2\big[c_n (k/n-\tau_j)\big]\Big)^2\no\\
&& +8\,\big( \frac {c_n}n\big)^2  \sum_{k=1}^n  \Big(\frac{1}{l_n}\sum_{1\le i<j\le l_n}K\big[c_n (k/n-\tau_i)\big]K\big[c_n (k/n-\tau_j)\big]\Big)^2\no\\
&\le &2\,C^2 \big( \frac {c_n}n\big)^2  \sum_{k=1}^n  \Big(\frac{1}{\sqrt {l_n}}\sum_{j=1}^{l_n}K\big[c_n (k/n-\tau_j)\big]\Big)^2+ 8I_n^2\to 0,
\eestar
due to (\ref {mai9a}) and (\ref {mai9a1}).
 $\Box$

\subsection{Proof of (\ref{double1})} \label{sect6.9}

Using similar arguments as in the proof of   (\ref {ad1}) or (\ref {sa2}), it suffices to show that
\[
\widetilde{I}_{n}:=\frac{c_{n}}{n\sqrt{l_{n}}}\sum_{k=1}^{n}\sum_{j=1}^{l_{n}}K\left[c_{n}\left(k/n-\tau_{j}\right)\right]K^{\ast}
\left[c_{n}\left(k/n-\tau^{\ast}\right)\right]\rightarrow 0.
\]
We first assume that $l_n\rightarrow\infty$.
For any $n\in \mathbb{N}$, there exists $i_n\in \mathbb{N}$ so that $i_n-1<\tau^{\ast}\le i_n$. Therefore, for any $j\ne i_n+1, i_n, i_n-1$,
we have $|\tau_{j}-\tau^{\ast}|\ge c_n(|j-i_n|-1)/(2l_n)$. This implies that $|k/n-\tau_j|\ge c_n(|j-i_n|-1)/(4l_n)$ or $|k/n-\tau^{\ast}|\ge c_n(|j-i_n|-1)/(4l_n)$.
Recall that $K(x)\leq C/(1+|x|)$ and  $K^*(x)\leq C/(1+|x|)$, we have
\begin{eqnarray*}
&&K\left[c_{n}\left(k/n-\tau_{j}\right)\right]K^{\ast}
\left[c_{n}\left(k/n-\tau^{\ast}\right)\right] \\
&&~~~~\le \frac{Cl_n}{c_n(|j-i_n|-1)}\Big(K\left[c_{n}\left(k/n-\tau_{j}\right)\right]+K^{\ast}
\left[c_{n}\left(k/n-\tau^{\ast}\right)\right]\Big).
\end{eqnarray*}
Therefore, by noting that $l_n=o(c_n)$ and $l_n\rightarrow \infty$,
\begin{eqnarray*}
\widetilde{I}_{n}&\le&\frac{C\sqrt{l_n}}{c_n}\sum_{|j-i_n|\ge 2} \frac{1}{|j-i_n|-1} \frac{c_{n}}{n}\sum_{k=1}^{n}
\Big(K\left[c_{n}\left(k/n-\tau_{j}\right)\right]+K^{\ast}
\left[c_{n}\left(k/n-\tau^{\ast}\right)\right]\Big)\\
&&~~~~+C\frac{c_{n}}{n\sqrt{l_{n}}}\sum_{k=1}^{n}K^{\ast}
\left[c_{n}\left(k/n-\tau^{\ast}\right)\right]\\
&\le& C\frac{\sqrt{l_n}\log l_n}{c_n}+C/\sqrt{l_n}\rightarrow 0.\end{eqnarray*}

Next, we assume that $l_n=l$ and $\tau^*, \tau_j, j=1,\dots, l$ are fixed constants.
If $\tau^{\ast}\neq \tau_j$,  then
\begin{eqnarray*}
&&\frac{c_{n}}{n}\sum_{k=1}^{n}K\left[c_{n}\left(k/n-\tau_{j}\right)\right]K^{\ast}
\left[c_{n}\left(k/n-\tau^{\ast}\right)\right]\\
&&~~~~\le \frac{2}{c_n|\tau_j-\tau^{\ast}|} \frac{c_{n}}{n}\sum_{k=1}^{n}\Big( K\left[c_{n}\left(k/n-\tau_{j}\right)\right]+K^{\ast}
\left[c_{n}\left(k/n-\tau^{\ast}\right)\right]\Big)\\
&&~~~~\le \frac{C}{c_n|\tau_j-\tau^{\ast}|} \rightarrow 0.
\end{eqnarray*}
This implies $\widetilde{I}_n\rightarrow 0$ for $\tau^*\not \in \{\tau_j, j=1,\dots, l\}$.
$\Box$

\bigskip{}

\end{document}